\pdfoutput=1
\documentclass[11pt,twoside,a4paper,cmspaper,final,collab]{cms-tdr}
\def\svnVersion{686fdc3}\def\svnDate{2026/01/13}\def\cmsCernNoTag{CERN-EP-2024-275}\def\cmsCernDate{\today}\def\cmsMessage{Published in the Journal of High Energy Physics as \href{http://dx.doi.org/10.1007/JHEP12(2025)052}{\doi{10.1007/JHEP12(2025)052}.}}
\begin{document}\cmsNoteHeader{EXO-22-013}

\DeclareRobustCommand{\PS}{{\HepParticle{S}{}{}}\Xspace}
\DeclareRobustCommand{\PSeu}{{\HepParticle{S}{\Pe\PQu}{}}\Xspace}
\DeclareRobustCommand{\PSed}{{\HepParticle{S}{\Pe\PQd}{}}\Xspace}
\DeclareRobustCommand{\PSmu}{{\HepParticle{S}{\PGm\PQu}{}}\Xspace}
\DeclareRobustCommand{\PSmd}{{\HepParticle{S}{\PGm\PQd}{}}\Xspace}

\DeclareRobustCommand{\PV}{{\HepParticle{V}{}{}}\Xspace}
\DeclareRobustCommand{\PVeu}{{\HepParticle{V}{\Pe\PQu}{}}\Xspace}
\DeclareRobustCommand{\PVed}{{\HepParticle{V}{\Pe\PQd}{}}\Xspace}
\DeclareRobustCommand{\PVmu}{{\HepParticle{V}{\PGm\PQu}{}}\Xspace}
\DeclareRobustCommand{\PVmd}{{\HepParticle{V}{\PGm\PQd}{}}\Xspace}

\newcommand{\fxfx}{{\textsc{FxFx}}\xspace}
\newcommand{\costhetastar}{\ensuremath{\cos\theta^*}\xspace}
\newcommand{\costhetar}{\ensuremath{\cos\theta_\mathrm{R}}\xspace}
\newcommand{\csr}{\ensuremath{c_\mathrm{R}}\xspace}
\newcommand{\cstar}{\ensuremath{c_*}\xspace}

\newcommand{\cmsTable}[1]{\resizebox{\textwidth}{!}{#1}}
\newcommand{\Zg}{\ensuremath{\PZ/\PGg^{*}}\xspace}
\newcommand{\glq}{\ensuremath{g_{\text{LQ}}}\xspace}
\newcommand{\ylq}{\ensuremath{y_{\text{LQ}}}\xspace}
\newcommand{\mlq}{\ensuremath{m_{\text{LQ}}}\xspace}
\newcommand{\mll}{\ensuremath{m_{\Pell\Pell}}\xspace}
\newlength\cmsTabSkip\setlength{\cmsTabSkip}{1ex}

\newcommand{\Rtwo}{\ensuremath{R_2}\xspace}
\newcommand{\Rttwo}{\ensuremath{\widetilde{R}_2}\xspace}
\newcommand{\Uthree}{\ensuremath{U_3}\xspace}
\newcommand{\yAB}[2]{\ensuremath{y_{#1#2}}\xspace}
\newcommand{\ylu}{\yAB{\Pell}{\PQu}}
\newcommand{\yld}{\yAB{\Pell}{\PQd}}
\newcommand{\ymu}{\yAB{\PGm}{\PQu}}
\newcommand{\yeu}{\yAB{\Pe}{\PQu}}
\newcommand{\ymd}{\yAB{\PGm}{\PQd}}
\newcommand{\yed}{\yAB{\Pe}{\PQd}}
\newcommand{\gAB}[2]{\ensuremath{g_{#1#2}}\xspace}
\newcommand{\glu}{\gAB{\Pell}{\PQu}}
\newcommand{\gld}{\gAB{\Pell}{\PQd}}
\newcommand{\gmu}{\gAB{\PGm}{\PQu}}
\newcommand{\geu}{\gAB{\Pe}{\PQu}}
\newcommand{\gmd}{\gAB{\PGm}{\PQd}}
\newcommand{\ged}{\gAB{\Pe}{\PQd}}
\newcommand{\Azero}{\ensuremath{A_0}\xspace}
\newcommand{\Afour}{\ensuremath{A_4}\xspace}
\newcommand{\Npure}{\ensuremath{N_\text{pure}}\xspace}
\newcommand{\Nint}{\ensuremath{N_\text{int}}\xspace}
\newcommand{\abseta}{\ensuremath{\abs{\eta}}\xspace}
\newcommand{\tautau}{\ensuremath{\PGt \PGt}\xspace}
\newcommand{\emu}{\ensuremath{\Pe\PGm}\xspace}
\newcommand{\tW}{\ensuremath{\PQt\PW}\xspace}
\newcommand{\ee}{\ensuremath{\Pe\Pe}\xspace}
\newcommand{\mumu}{\ensuremath{\Pgm\Pgm}\xspace}
\newcommand{\fs}{\ensuremath{f_\mathrm{s}}\xspace}
\newcommand{\fa}{\ensuremath{f_\mathrm{a}}\xspace}
\newcommand{\falpha}{\ensuremath{f_{\alpha}}\xspace}
\newcommand{\flqp}{\ensuremath{f_\text{LQ(pure)}}\xspace}
\newcommand{\flqi}{\ensuremath{f_\text{LQ(int)}}\xspace}

\cmsNoteHeader{EXO-22-013}

\title{Search for \texorpdfstring{$t$}{t}-channel scalar and vector leptoquark exchange in the high-mass dimuon and dielectron spectra in proton-proton collisions at \texorpdfstring{$\sqrt{s} = 13\TeV$}{sqrt(s) = 13 TeV}}

\date{\today}

\abstract{
  A search for $t$-channel exchange of leptoquarks (LQs) is performed in dimuon and dielectron spectra using proton-proton collision data collected at $\sqrt{s}=13\TeV$ with the CMS detector at the CERN LHC. The data correspond to an integrated luminosity of 138\fbinv. Eight scenarios are considered, in which up or down quarks couple to muons or electrons via a scalar or vector LQ exchange, for dilepton invariant masses above 500\GeV. The LQ masses are probed up to 5\TeV, beyond a regime probed by previous pair-production and single-production searches. The differential distributions of dilepton events are fit to templates that model the nonresonant LQ exchange and various standard model background processes. Limits are set on LQ-fermion coupling strengths for scalar and vector LQ masses in the 1--5\TeV range at $95\%$ confidence level, establishing stringent limits on first- and second-generation LQs. 
}

\hypersetup{
pdfauthor={CMS Collaboration},
pdftitle={Search for t-channel scalar and vector leptoquark exchange in the high mass dimuon and dielectron spectrum in proton-proton collisions at sqrt(s) = 13 TeV},
pdfsubject={CMS},
pdfkeywords={LQs, CMS, Drell-Yan, scalar, vector}}

\maketitle

\section{Introduction}
Leptoquarks (LQs) are hypothetical particles proposed to explain the similarities between the quark and lepton sectors of the standard model (SM). Leptoquarks were initially introduced in early extensions of the SM, for example, in grand unified theories based on the $SU(5)$ and $SO(10)$ groups~\cite{Pati:1974yy,Georgi:1974sy,Senjanovic:1982ex,Davidson:2010uu,Dorsner:2005fq,Buchmuller:1986iq}, such as the Pati--Salam model~\cite{Pati:1974yy}, composite models~\cite{Gripaios:2014tna,Gripaios:2009dq}, and technicolor models~\cite{Farhi:1980xs,Dimopoulos:1979sp,Schrempp:1984nj}. 
Leptoquarks have also been proposed as mediators of interactions between SM and dark matter particles~\cite{Baker:2015qna,Choi:2018stw}, and appear in $R$-parity violating supersymmetry models~\cite{Barbier:2004ez}. 
  Results from the Muon $g-2$ experiment in 2021 showed deviations in the experimental value of the anomalous magnetic moment of the muon to 0.46 ppm~\cite{Muong-2:2021ojo}. The deviation is reduced when updated lattice quantum chromodynamics (QCD) calculations are considered~\cite{Borsanyi:2020mff}. 
Although the tension of the $g-2$ measurements has been largely resolved by using these calculations, a massive LQ could contribute to the anomalous magnetic moment by mixing with the SM Higgs field~\cite{Dorsner:2019itg}, and thus move the theory even closer to the latest experimental result~\cite{Muong-2:2025xyk}. 

Leptoquarks are electrically charged, color-triplet bosons, either scalar or vector, that carry a baryon number ($B$), as well as a lepton number ($L$). 
They have a fermion number $F = 3B+L$ of 0 or $-2$; however, LQs with $F=-2$ can mediate proton decay, and stringent limits from dedicated proton decay searches have strongly constrained such LQs~\cite{Dorsner:2016wpm}. 
Distinctions between different LQs are based on their SM gauge group transformation properties, giving rise to twelve LQ multiplets. 
Only six families do not mediate proton decay. 
Leptoquarks are labeled as first-, second-, third-, or mixed-generation based on the SM quarks and leptons with which they interact.  

\begin{figure}[hbtp]
  \centering
   \includegraphics[width=0.7\textwidth]{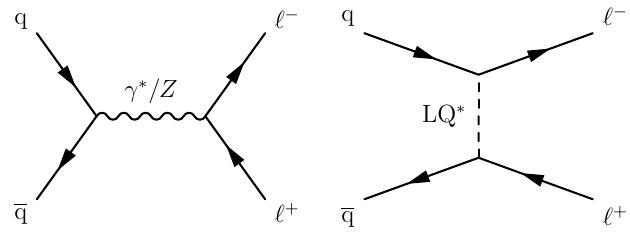}
  \caption{
  Leading order Feynman diagrams for SM DY production (left) and $t$-channel LQ exchange (right). 
  The LQ amplitude interferes with the \Zg amplitude.}
  \label{fig:DY}
\end{figure}

The quark-LQ-lepton interaction is characterized by the LQ-fermion coupling constants \ylq (\glq) for the scalar (vector) case.
At colliders, LQs can be produced singly or in pairs, or manifest via $t$-channel exchange. 
The pair-production channel is dominated by gluon-gluon fusion, with a cross section proportional to the square of the strong coupling constant $(\alpS^2)$.
Since this production mode proceeds via the strong interaction, it is insensitive to the LQ-fermion coupling, and results in the most stringent bounds at low LQ masses ($\mlq\sim 1\TeV$). 
The single-production cross section scales as $\alpS\ylq^2$ ($\alpS\glq^2$), but the channel can be more sensitive than pair production at higher $\mlq$ or $\ylq$ ($\glq$). 
Finally, the $t$-channel exchange process, shown in Fig.~\ref{fig:DY}, has a cross section proportional to $\ylq^4$ ($\glq^4$), with additional contributions proportional to $\ylq^2$ ($\glq^2$) from the interference with dilepton production via the Drell--Yan (DY) process. 
This channel can have superior sensitivity at even higher \mlq, as its cross section decreases more slowly than for the other two channels, as a function of \mlq. 

Several searches for LQs have been performed by the ATLAS and CMS Collaborations. 
Third-generation scalar and vector LQs coupled to a tau (\PGt) lepton and a bottom (\PQb) quark were studied in Refs.~\cite{CMS:2023qdw,ATLAS:2023uox}, which considered single and pair production, as well as $t$-channel LQ exchange.
The $t$-channel exchange process is the most sensitive channel at large \mlq and \ylq (\glq), excluding at 95\% confidence level (\CL) third-generation scalar (vector) LQs with masses of 0.5--1.5 (0.5--2.3)\TeV for \ylq (\glq) values larger than 1.8--3.5 (0.9--3.5). Searches for vector LQs in the $\tautau$ channel were reported in Ref.~\cite{CMS:2023arc}, in which \mlq of 1--5\TeV were excluded for $\glq > 0.8$--$3.7$ at 95\% \CL for a $U_1$ LQ in the Pati--Salam model.  

For pair-produced scalar LQs coupled to first- and second-generation fermions, the most stringent limits have been set by ATLAS, excluding LQs with masses below 1.8 and 1.7\TeV at 95\% \CL, respectively, independent of $\ylq$ and assuming a branching fraction of 100\% into a quark and a lepton~\cite{ATLAS:2020dsk}.
Single production of first- and second-generation scalar LQs has been studied by CMS~\cite{CMS:2015xzc}, in which \mlq below 1.73\TeV and 530\GeV, respectively, were excluded, assuming $\glq=1$ and a branching fraction of 100\% into a quark and a lepton. 
Pair production of first- and second-generation scalar LQs have also been studied by CMS~\cite{CMS:2018ncu,CMS:2018lab}. 
First-generation scalar \mlq were excluded up to 1.44\TeV at 95\% \CL, and second-generation scalar \mlq were excluded up to 1.53\TeV, independent of \ylq and assuming a branching fraction of 100\% into a quark and a lepton. The CMS Collaboration has also studied pair production of first- and second-generation vector LQs, leading to exclusions of \mlq up to 1.5 (1.3)\TeV assuming Yang--Mills (minimal) couplings, and \mlq less than 1.7 (1.1)\TeV assuming minimal-minimal (absolute minimal) couplings, for a branching fraction of 100\% into a quark and a lepton~\cite{CMS:2015gua}.

Motivated by the phenomenological studies performed in Refs.~\cite{Raj:2016aky, Crivellin:2021egp}, three families of LQs are considered: two scalar doublet families, \Rtwo and \Rttwo, and one vector triplet family, \Uthree. 
The subscripts refer to the dimensionality of the $SU(2)$ representation. 
Within the \Rtwo, \Rttwo, and \Uthree families, first- and mixed-generation LQs that couple up and down quarks to electrons and muons are considered. 
For the scalar doublet families, following Ref.~\cite{Raj:2016aky}, the case where the LQ couples only to right-handed quark multiplets and left-handed lepton multiplets (RL) is considered. 
The LQs are assumed to couple to a single quark and lepton, resulting in four scalar LQ coupling scenarios: two \Rtwo LQs, \PSeu and \PSmu, and two \Rttwo LQs, \PSed and \PSmd, where the subscripts denote the SM fermions. 
Similarly, four \Uthree vector LQs are considered: \PVeu, \PVmu, \PVed, and \PVmd. 
The SM gauge group transformation properties for the chosen LQ models are summarized in Table~\ref{tab:lq_props}. 
The \Rtwo LQs appear in potential solutions to the muon anomalous magnetic moment problem~\cite{Bigaran:2020jil_r2,Dorsner:2023ykm}, \Rttwo LQs appear in beyond-the-SM extensions for lepton flavor universality violation involving sterile neutrinos~\cite{Robinson:2018gza,Dorsner:2023ykm}, and \Uthree LQs appear in models that describe scenarios for lepton flavor violation~\cite{Becirevic:2016oho_u3}.

\begin{table}[!htbp]
  \centering
  \topcaption{
  Properties of the \Rtwo, \Rttwo, and \Uthree LQs from Ref.~\cite{Dorsner:2016wpm}. 
  This paper describes a search for \Rtwo LQs with RL couplings and charge 5/3, \Rttwo LQs with RL couplings and charge 2/3, and \Uthree LQs with charges 2/3 and 5/3.}
  \label{tab:lq_props}
  \renewcommand{\arraystretch}{1.3}
  \begin{tabular}{lccc}
    LQ family & $(SU(3),\,SU(2),\,U(1))$ & Spin & Charge\\
    \hline
     \Rtwo & (3, 2, 7/6) & 0 & 5/3, 2/3\\
     \Rttwo & (3, 2, 1/6) & 0 & 2/3, -1/3\\
     \Uthree & (3, 3, 2/3) & 1  & 5/3, 2/3, -1/3 \\
  \end{tabular}
\end{table}

The LQ-quark-lepton interactions are defined by the following terms in the Lagrangian:
\begin{equation}
\mathcal{L} \supset \ylu\overline{\PQu}_R \Pell_L R^{5/3} + \yld\overline{\PQd}_R \Pell_L \widetilde{R}^{2/3} + \glu\overline{\PQu}_L\gamma^\mu\Pell_L U^{5/3}_\mu + \gld\overline{\PQd}_L\gamma^\mu \Pell_L U^{2/3}_\mu  + \mathrm{h.c.},
\end{equation}
where \ylu (\yld) are the couplings of the charge 5/3 (2/3) \Rtwo (\Rttwo) scalar LQs to fermions and \glu (\gld) are the couplings of the charge 5/3 (2/3) vector \Uthree LQs to fermions. The terms describing couplings of LQs to neutrinos are omitted. A complete description of the LQ interactions, as well as a more detailed review of LQ physics in precision experiments, can be found in Ref.~\cite{Dorsner:2016wpm}.

This paper describes the first search for the $t$-channel exchange of LQs coupled to first- and second-generation SM fermions, probing scalar and vector LQs with masses in the range 1--5\TeV.
The search uses proton-proton ($\Pp\Pp$) collision data at $\sqrt{s}=13\TeV$ corresponding to an integrated luminosity of 138\fbinv, collected by the CMS experiment at the CERN LHC in 2016--2018.
The hypothetical signal process results in the nonresonant exchange of muon or electron pairs. It is distinguished from the dominant background of SM DY events using the differential distributions of the dilepton invariant mass, the angle between the incoming quark and the outgoing lepton, and the rapidity of the dilepton system. 
The observed dilepton distributions are fit with a combination of signal and background templates, both produced by reweighting simulated SM DY events. 

The paper is divided into ten sections. The analysis strategy is discussed in Section~\ref{sec:analysis_strategy}, followed by a description of the CMS detector and event reconstruction in Section~\ref{sec:detector}. The simulated samples used for the signal and background estimation are described in Section~\ref{sec:data_mc_samples}, and the criteria used to select dimuon and dielectron events are described in Section~\ref{sec:event_selection}. Section~\ref{sec:backgrounds} contains a detailed discussion of the various backgrounds for this search, and how they are estimated. The procedure to build templates of the signal and background distributions is described in Section~\ref{sec:temp_construct}, followed by a discussion of the various systematic uncertainties that affect the measurements in Section~\ref{sec:sys_uncs}. The results of maximum likelihood fits to data are discussed in Section~\ref{sec:results}, along with the limits at 95\% \CL on eight LQ couplings as functions of \mlq. Finally, the paper is summarized in Section~\ref{sec:summary}.

Tabulated results are provided in the HEPData record for this analysis~\cite{HEPData}.

\section{Analysis strategy}\label{sec:analysis_strategy}

The $t$-channel LQ exchange occurs in conjunction with the SM DY process, which is the primary background for this search. 
This analysis utilizes three variables computed from the dilepton system: the invariant mass, \mll; the rapidity $y$; and \costhetastar (abbreviated as \cstar), the cosine of the angle between the quark (as opposed to the antiquark) and the lepton (as opposed to the antilepton) in the Collins--Soper frame (dilepton rest frame)~\cite{Collins:1977iv,CMS:2022uul}. 
At a $\Pp\Pp$ collider, the direction of the incident quark is unknown. Instead, an approximation is used, defined as $\costhetar \equiv \frac{p_z}{\abs{p_z}} \cstar$, abbreviated as \csr, where $p_z$ stands for the longitudinal momentum of the dilepton system. This approximation uses the fact that the quark will typically have higher longitudinal momentum than the antiquark, and hence, the longitudinal momentum of the dilepton system will more often align with that of the quark.
The observed distributions in these three variables can be expressed as a convolution of the underlying differential cross section:
\begin{equation}
\label{eq:ang_xsec2}
  f(\mll^\text{r},\,y^\text{r},\,\csr) = C \int \rd \mll\, \rd y\, \rd \cstar R(\mll^\text{r},\,y^\text{r},\,\csr;\,\mll,\,y,\,\cstar)
  \varepsilon (\mll,\,y,\,\cstar) \frac{\rd^3\sigma}{\rd \mll\, \rd y\, \rd \cstar} ,
\end{equation}
where the superscript \text{r} distinguishes reconstructed variables from generator-level variables, $C$ is a normalization constant, $R$ is a resolution function that incorporates detector resolution and parton shower effects, 
and $\varepsilon$ is an efficiency function.  
Since the convolution is a linear function of the cross section, $\sigma$, the resulting reconstructed distribution can be expressed as a sum of convolutions of different contributions to the total cross section. 
The linearity allows the expected observation for a given signal hypothesis to be represented as a sum of templates that are individually independent of the parameters of interest. 
Specifically, the parton-level differential cross section of the SM DY process can be written as:
\begin{equation}
  \label{eq:ang_xsec}
  \left[\frac{\rd\sigma}{\rd \cstar}(\mll^2)\right]_\mathrm{DY} \propto \frac{3}{8}\biggl\lbrace (1 + \cstar^2) + \Afour\cstar + \frac{\Azero}{2} \left(1-3 \cstar^2\right) \biggr\rbrace,
\end{equation}
where \Afour and \Azero are the standard dimensionless coefficients parametrizing the angular distribution of the DY process ~\cite{Mirkes:1992hu,Mirkes:1994dp}. 
The first two terms correspond to the leading order (LO), and the third term corresponds to next-to-LO (NLO). 
The coefficients \Azero and \Afour are functions of \mll, the transverse momentum (\pt), and $y$ of the dilepton system. 
Measurements of \Azero probe higher-order corrections in perturbative QCD, whereas \Afour is directly related to the SM forward-backward asymmetry, $\AFB = (\sigma_\text{F} - \sigma_\text{B}) / (\sigma_\text{F} + \sigma_\text{B})=\frac{3}{8}\Afour$, where $\sigma_\text{F}$ and $\sigma_\text{B}$ are the cross sections for forward ($\cstar>0$) and backward ($\cstar<0$) events, respectively. A measurement of \AFB and \Azero was recently performed by CMS~\cite{CMS:2022uul}.

The LO double-differential cross section for DY production that includes the $t$-channel exchange of a scalar LQ can be expressed as~\cite{Raj:2016aky}:
\begin{equation}
  \label{eq:ang_xsec_reparam_LQ}
  \begin{aligned}
    \frac{\rd^2\sigma}{\rd \mll\rd \cstar} \propto& \left[\frac{\rd^2 \sigma}{\rd \mll \rd \cstar}\right]_\mathrm{DY}  + \ylq^4 \Npure^S(\mll)\left(\frac{1-\cstar}{1-\cstar+\frac{2\mlq^2}{\mll^2}}\right)^2 \\
    &+\, \ylq^2 \Nint^S(\mll)\left[\frac{(1-\cstar)^2}{1-\cstar+\frac{2\mlq^2}{\mll^2}}\right],
  \end{aligned}
\end{equation}
and the differential cross section for a vector LQ exchange can be expressed as:
\begin{equation}
  \label{eq:ang_xsec_reparam_LQ_vec}
  \begin{aligned}
  \frac{\rd^2\sigma}{\rd \mll\rd \cstar} \propto& \left[\frac{\rd^2 \sigma}{\rd \mll \rd \cstar}\right]_\mathrm{DY}  + \glq^4 \Npure^V(\mll)\left(\frac{1+\cstar}{1-\cstar+\frac{2\mlq^2}{\mll^2}}\right)^2 \\
  &+\, \glq ^2 \Nint^V(\mll) \left[\frac{(1+\cstar)^2}{1-\cstar+\frac{2\mlq^2}{\mll^2}}\right],
\end{aligned}
\end{equation}
where $\Npure^{S,V}$ and $\Nint^{S,V}$ are prefactors that depend on \mll, and the vector and axial-vector couplings of the quark and lepton that couple to the LQ. 
The first LQ-specific term corresponds to pure $t$-channel LQ exchange, while the second term corresponds to the interference of the $t$-channel LQ exchange with the SM \Zg diagram.  

Following an approach similar to that of Ref.~\cite{CMS:2022uul}, the templates are histograms binned in three variables (\mll, \csr, and $y$), constructed from simulated SM DY events that are reweighted using functions of the generator-level variables to represent each of the terms in Eqs.~\eqref{eq:ang_xsec_reparam_LQ} or \eqref{eq:ang_xsec_reparam_LQ_vec}. 
A maximum likelihood fit of a linear combination of the templates to the observed data is performed to extract the signal and background parameters, namely \Azero, \Afour, and either \ylq or \glq. 
The template construction is described in more detail in Section~\ref{sec:temp_construct}. 
Each of the eight combinations of spin and coupling to SM fermions is probed separately. 

\section{The CMS detector and event reconstruction} \label{sec:detector}
The central feature of the CMS apparatus is a superconducting solenoid of 6\unit{m} internal diameter, providing a magnetic field of 3.8\unit{T}. 
Within the solenoid volume are a silicon pixel and strip tracker, a lead tungstate crystal electromagnetic calorimeter (ECAL), and a brass and scintillator hadron calorimeter (HCAL), 
each composed of a barrel and two endcap sections. 
Forward calorimeters extend the pseudorapidity ($\eta$) coverage provided by the barrel and endcap detectors. 
Muons are detected in gas-ionization chambers embedded in the steel flux-return yoke outside the solenoid.
A more detailed description of the CMS detector, together with a definition of the coordinate system used and the relevant kinematic variables, is reported in Ref.~\cite{CMS:2008xjf,CMS:2023gfb}.

Events of interest are selected using a two-tiered trigger system. 
The first level, composed of custom hardware processors, uses information from the calorimeters and muon detectors 
to select events at a rate of around 100\unit{kHz} within a fixed latency of about 4\mus~\cite{Sirunyan:2020zal}. 
The second level, known as the high-level trigger, consists of a farm of processors running a version of the full event reconstruction software 
optimized for fast processing, and reduces the event rate to around 1\unit{kHz} before data storage~\cite{Khachatryan:2016bia}.

A particle-flow algorithm~\cite{Sirunyan:2017ulk} aims to reconstruct and identify each individual particle in an event, 
with an optimized combination of information from the various elements of the CMS detector. 
The energy of photons is obtained from the ECAL measurement. 
The energy of electrons is determined from a combination of the electron momentum at the primary interaction vertex as determined by the tracker, 
the energy of the corresponding ECAL cluster, and the energy sum of all bremsstrahlung photons spatially compatible with originating from the electron track. 
The energy of muons is obtained from the curvature of the corresponding track. 
The energy of charged hadrons is determined from a combination of their momentum measured in the tracker and the matching ECAL and HCAL energy deposits, 
corrected for the response function of the calorimeters to hadronic showers. 
Finally, the energy of neutral hadrons is obtained from the corresponding corrected ECAL and HCAL energies.
The primary vertex is taken to be the vertex corresponding to the hardest scattering in the event, evaluated using tracking information alone, as described in Section 9.4.1 of Ref.~\cite{CMS-TDR-15-02}.

Muons are measured in the range $\abseta<2.4$, with detection planes made using three technologies: drift tubes, cathode strip chambers, and resistive-plate chambers. 
The single-muon trigger efficiency exceeds 90\% over the full $\eta$ range, and the efficiency to reconstruct and identify muons is greater than 96\%. 
Matching muons to tracks measured in the silicon tracker results in a relative \pt resolution of 1\% in the barrel and 3\% in the endcaps for muons with $\pt<100\GeV$,
and of better than 7\% for muons in the barrel with $\pt<1\TeV$~\cite{CMS:2018rym}.

The single-electron trigger efficiency is approximately 80\% over the full $\eta$ range, 
and the efficiency to reconstruct and identify electrons is greater than 65\% for electrons with $\pt>20\GeV$. 
The momentum resolution for electrons with $\pt\approx 45\GeV$ from $\PZ \to \ee$ decays ranges from 1.7--4.5\%. 
It is generally better in the barrel region than in the endcaps, and also depends on the bremsstrahlung energy emitted by the electron as it traverses the material in front of the ECAL~\cite{CMS:2020uim}.

Additional $\Pp\Pp$ interactions within the same or nearby bunch crossings (pileup) 
can contribute additional tracks and calorimetric energy depositions, increasing the apparent jet momentum. 
To mitigate this effect, tracks identified as originating from pileup vertices are discarded and an offset correction is applied to correct for remaining contributions~\cite{Sirunyan_2020pileup}. 

\section{Simulated event samples} \label{sec:data_mc_samples}

Various Monte Carlo (MC) event generators are used to simulate the SM background processes. 
The DY samples are generated at NLO in QCD with up to two extra partons using \MGvATNLO \cite{Alwall:2014hca} v2.2.2 (v2.6.0) for samples corresponding to the 2016 (2017--2018) data-taking periods.
The parton shower, hadronization, and quantum electrodynamics final-state radiation are simulated with \PYTHIA~v8.226 (v8.230)~\cite{Sjostrand:2014zea}, using the \fxfx merging scheme and the underlying event tune CUETP8M1 (CP5)~\cite{CMS:2019csb,CMS:2015wcf}.
The NNPDF 3.0 parton distribution functions (PDFs)~\cite{Ball:2010de, Ball:2014uwa} at NLO accuracy are used for samples corresponding to 2016, and NNPDF 3.1~\cite{Ball:2018iqk} at next-to-NLO (NNLO) are used for 2017--2018. For consistency, all samples were reweighted to NNPDF 3.0 at NLO.

Additional backgrounds include diboson production 
($\PW\PW$, $\PW\PZ$, $\PZ\PZ$), photon-induced dilepton production ($\PGg \PGg \to \Pell \Pell$), 
top quark pair production (\ttbar), and single top quark production in association with a $\PW$ boson (\tW).
The \ttbar and \tW backgrounds are generated at NLO using \POWHEG v2.0~\cite{Nason:2004rx,Frixione:2007vw,Alioli:2010xd,Frixione:2007nw}, 
and interfaced with \PYTHIA~v8.226 (v8.230), with 
the CUETP8M2T4 (CP5) underlying event tune~\cite{CMS-PAS-TOP-16-021}  for the 2016 (2017--2018) data-taking period.
The diboson background samples are generated at NLO, with $\PZ\PZ \to \Pell \Pell \Pell^{\prime} \Pell^{\prime}$, $\PZ\PZ \to \Pell \Pell \PGn \PGn$, and $\PW\PW \to \Pell \PGn \Pell \PGn$ processes generated using \POWHEG~\cite{Melia:2011tj,Nason:2013ydw} and  $\PW\PZ \to \PQq\PQq \Pell \Pell$, $\PW\PZ \to \Pell \Pell \Pell\PGn$, and $\PZ\PZ \to \PQq\PQq \Pell \Pell$ processes generated using \MGvATNLO. 
All diboson samples are showered using \PYTHIA and tune CUETP8M1 (CP5) for the 2016 (2017--2018) data-taking periods. 
The \ttbar, \tW, and diboson backgrounds use the NNPDF 3.0 (3.1) PDFs for 2016 (2017--2018) data-taking periods. 
The photon-induced background,
$\PGg \PGg \to \Pell \Pell$, is simulated using the \textsc{CepGen}~\cite{Forthomme:2018ecc} implementation of \textsc{LPair}~\cite{Vermaseren:1982cz,Baranov:1991yq},
interfaced with \PYTHIA v6.429~\cite{Sjostrand:2006za}, using the Suri--Yennie proton structure function parametrization~\cite{Suri:1971yx}.
This contribution is split into three parts because the interaction at each proton vertex
can be either elastic or inelastic.

The cross sections for the $\PW\PZ$ and $\PZ\PZ$ samples are normalized to the NLO predictions calculated
with \textsc{mcfm}~6.6~\cite{Campbell:2015qma}, while the cross sections of the $\PW\PW$
samples are normalized to the next-to-next-to-leading-logarithmic accuracy predictions~\cite{Gehrmann:2014fva}. The total cross section for the \ttbar
process is normalized to the prediction with NNLO accuracy in QCD and next-to-next-to-leading-logarithmic accuracy for the soft-gluon radiation
resummation calculated with \TOPpp2.0~\cite{Czakon:2011xx}.

Leptoquark signal samples modeled in the \textsc{SLQrules} package~\cite{Crivellin:2021ejk} are generated using \MGvATNLO. 
These samples are used only for cross checks on the LQ template method. 

The detector response for all MC samples is simulated using \GEANTfour~\cite{GEANT4:2002zbu}. 
The pileup distribution in simulation is weighted to match the one observed in data.

\section{Event selection} \label{sec:event_selection}
Events are required to have two leptons of same flavor and opposite charges.
The dimuon and dielectron events are selected by single-muon and single-electron triggers, respectively. Dilepton triggers were not used as they had negligible gains over using single-lepton triggers. The events selected by the triggers are reconstructed and required to satisfy a number of
offline requirements.
The leading muon and electron \pt requirement is at least 40\GeV.
The subleading lepton is required to have $\pt>15\GeV$. 
All muons are required to be within the acceptance of the muon system ($\abseta<2.4$), and all electrons must be within $\abseta<2.5$, excluding the barrel-endcap transition
region of the ECAL ($1.44<\abseta<1.57$) because the reconstruction of an electron object in this region is not optimal.
Additionally, to remove cosmic ray muon induced events, the azimuthal angle ($\phi$) between the two muons is required to differ from $\pi$ by more than 5\unit{mrad}. 

Each reconstructed muon is required to pass identification criteria that are based on the number of hits observed in the tracker, the response of the muon detectors,
and a set of matching criteria between muon track parameters, as measured by the inner tracker and muon detectors. 
To suppress nonprompt muons coming from heavy-flavor decays, each muon must be isolated from other particles in a cone of size $\Delta R=0.4$ 
where $\Delta R = \sqrt{\smash[b](\Delta \eta)^2 + (\Delta \phi)^2}$ refers to the distance from the muon to a given track. Further, the sum of the energy deposits of the charged and neutral hadrons divided by the muon track \pt within this cone should be less than 0.15.
More details on the muon identification and reconstruction used in this analysis can be found in Refs.~\cite{CMS:2018rym,CMS:2019ied}. 

The reconstructed electron candidates are required to pass identification criteria that are based on electromagnetic shower shape variables. 
Electrons originating from photon conversions are suppressed by requiring that the candidates have at most one missing inner tracker hit 
and not be consistent with being part of a conversion pair. 
Electrons are also required to be isolated from other particles within a cone of $\Delta R=0.3$. 
The electron isolation criteria are based on the ratios of the electron \pt to the sum of energy deposits associated with the
photons or hadrons.
More details on the electron reconstruction and identification criteria used in this analysis
are described in Ref.~\cite{CMS:2020uim}.

Events in the signal region (SR) must have $\mll>500\GeV$. This requirement suppresses \ttbar and nonprompt backgrounds. Additionally, the requirement has negligible impact on the signal sensitivity for heavy LQs ($\mlq>1\TeV$), which primarily comes from events with $\mll > 1\TeV$.

\section{Backgrounds} \label{sec:backgrounds}

The dominant background comes from DY events, and the subdominant background consists of fully leptonic decays of \ttbar events. 
Backgrounds from \tW events, diboson processes, and leptonic decays of \tautau originating from the DY process are also included.
All of these backgrounds are modeled in simulation with a few corrections based on control samples in data. 
The estimated event yields of the \ttbar, \tW, \tautau, and diboson events are validated in data in a control region (CR) of \emu events. A small fraction of the background comes from $t$-channel photon-induced dilepton production, which is modeled with MC simulation as well. Additionally, events where one or more jets are incorrectly identified as leptons also contribute to the total background yield. 
Such events are referred to as ``MisID'' events, primarily arising from $\PW$+jets processes and events composed uniquely of jets produced through the strong interaction, referred to as QCD multijet events. A technique based on CRs in data is employed to estimate the MisID contribution. For the dimuon (dielectron) channel, the total background from simulation and CR-based estimation consists of approximately 85 (82)\% DY, 10 (9)\% \ttbar plus \tW, 5 (4)\% diboson, and 0.5 (5)\% MisID events. 

The MisID background is estimated from data using the ``misidentification rate'' method~\cite{CMS:2018mdl,CMS:2014lcz}.
The misidentification rate is defined as the probability of a jet, having been reconstructed as a lepton candidate,
to pass the lepton selection requirements outlined in Section~\ref{sec:event_selection}.
This rate is measured in a sample with
two reconstructed leptons from a \PZ boson decay and an additional, potentially misidentified lepton.  
The two leptons corresponding to a \PZ boson decay are required to pass the lepton identification requirements and 
have an invariant mass within 7\GeV of the \PZ boson mass~\cite{ParticleDataGroup:2024cfk}.
The third lepton candidate is used as a probe to measure the misidentification rate.
This third lepton candidate is required to pass a less stringent set of identification and isolation requirements
than the full selection requirements of the analysis.
The MisID background can then be estimated from CRs of data events with two lepton 
candidates: one region in which one lepton fails the full selection requirements, to model the $\PW$+jets events, and another region in which both leptons fail the requirements, to model the QCD multijet events. 
Events from these regions are assigned weights based on the expected misidentification probabilities of
the failing lepton candidates. 
Contamination of this region by lepton pairs from DY and other prompt processes is subtracted using MC simulation.
These reweighted events are used to estimate the yield and shape of the MisID background. 

The MisID background estimation is validated in a CR with the same selection criteria as the SR, 
except the lepton pairs are required to have same-sign (SS) rather than opposite-sign (OS) charges.
A large fraction of the events in this CR stem from misidentified jets.
In the \ee channel, there is a significant contribution to this sample from OS DY events where the charge of one of the electrons has been incorrectly assigned.
The rate of charge misassignment in simulation is corrected using a sample of dilepton events near the \PZ peak with $70<\mll<110\GeV$.
The MisID background estimate as well as MC estimates of other backgrounds are compared with the observed yield of SS events.
A \csr-dependent correction to the MisID background estimate is calculated using the ratio of the MisID estimate to the number of observed SS events minus other backgrounds.
The uncertainty in this correction is calculated as the quadratic sum of the statistical uncertainties from the limited sample size, the
uncertainty in the number of DY pairs reconstructed as SS pairs, 
as well as a systematic uncertainty reflecting possible differences in shape between SS and OS MisID estimates.

The \tautau, \tW, \ttbar, and diboson backgrounds are validated in data using \emu events. 
The \emu CR has the same selection criteria as the SR, except that events are required to have one muon and one electron
rather than pairs of same flavor. The CR is constructed from events that have a leading muon and a subleading electron, because samples with high-quality muons have lower contamination from misidentified leptons. 
The muon is required to have $\pt>40\GeV$ and pass the single-muon trigger, 
and the electron is required to have $\pt>15\GeV$. The identification and isolation criteria for muons and electrons are the same as in the SR.
The MisID background is also present in this \emu CR and is estimated using the misidentification rate technique previously described. 
Good agreement is observed between the simulated and observed yields of \emu events across the entire mass range.

Because diboson events are produced via electroweak processes, they are expected to have a small forward-backward asymmetry in their lepton 
pairs; \ttbar events are also known to have an asymmetry, but it is too small to be detected using leptons alone~\cite{Sirunyan:2019eyu}.
The QCD and $\PW$+jets backgrounds are expected to have no asymmetry. 
Based on MC and CR estimates, an overall $\AFB\approx0.01$ is predicted in
the region of \emu events. 
The observed value in the \emu CR is $\AFB=0.012\pm0.003$, consistent with this expectation.

Although the overall normalizations and asymmetries in \emu events are observed to be consistent between data and simulation, 
there are discrepancies between the predicted and observed shapes of the \csr distributions. 
To address these discrepancies, a correction is derived from the ratio of the \csr distributions in data and simulation in \emu events, and applied to the shape of the \csr distributions in simulation.
Because the asymmetry is modeled well, this correction is derived symmetrically in \csr using four bins of $\abs{\csr}$.
A single correction is derived in the 500--3000\GeV mass range because of the limited event count.
This correction modifies by 5--25\% the shapes of the templates used to model the corresponding backgrounds in the SR. The statistical uncertainty in each of the four $\abs{\csr}$ bins is modeled as a separate shape nuisance parameter.  

\begin{figure}[hbtp]
  \centering
  \includegraphics[width = 0.44 \textwidth]{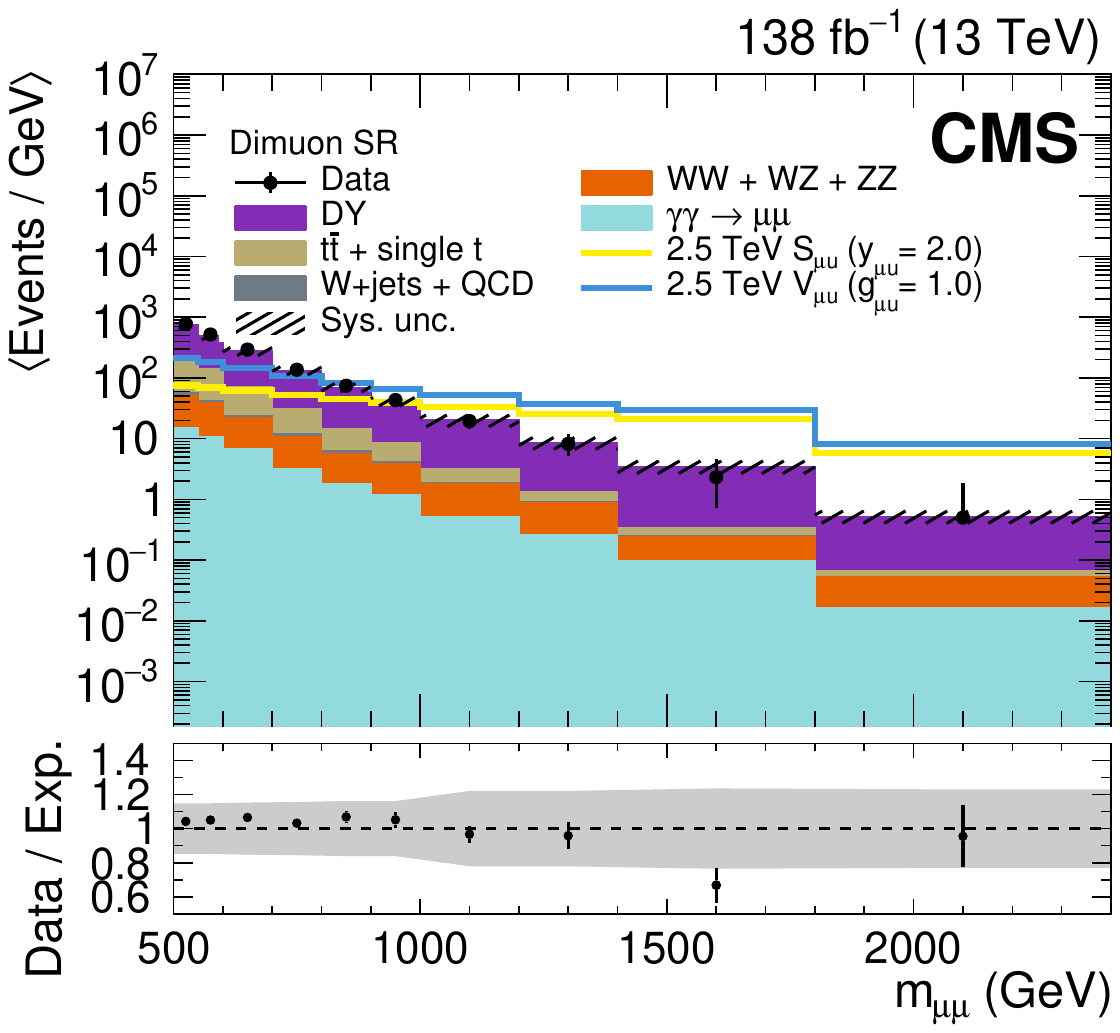}
  \includegraphics[width = 0.44 \textwidth]{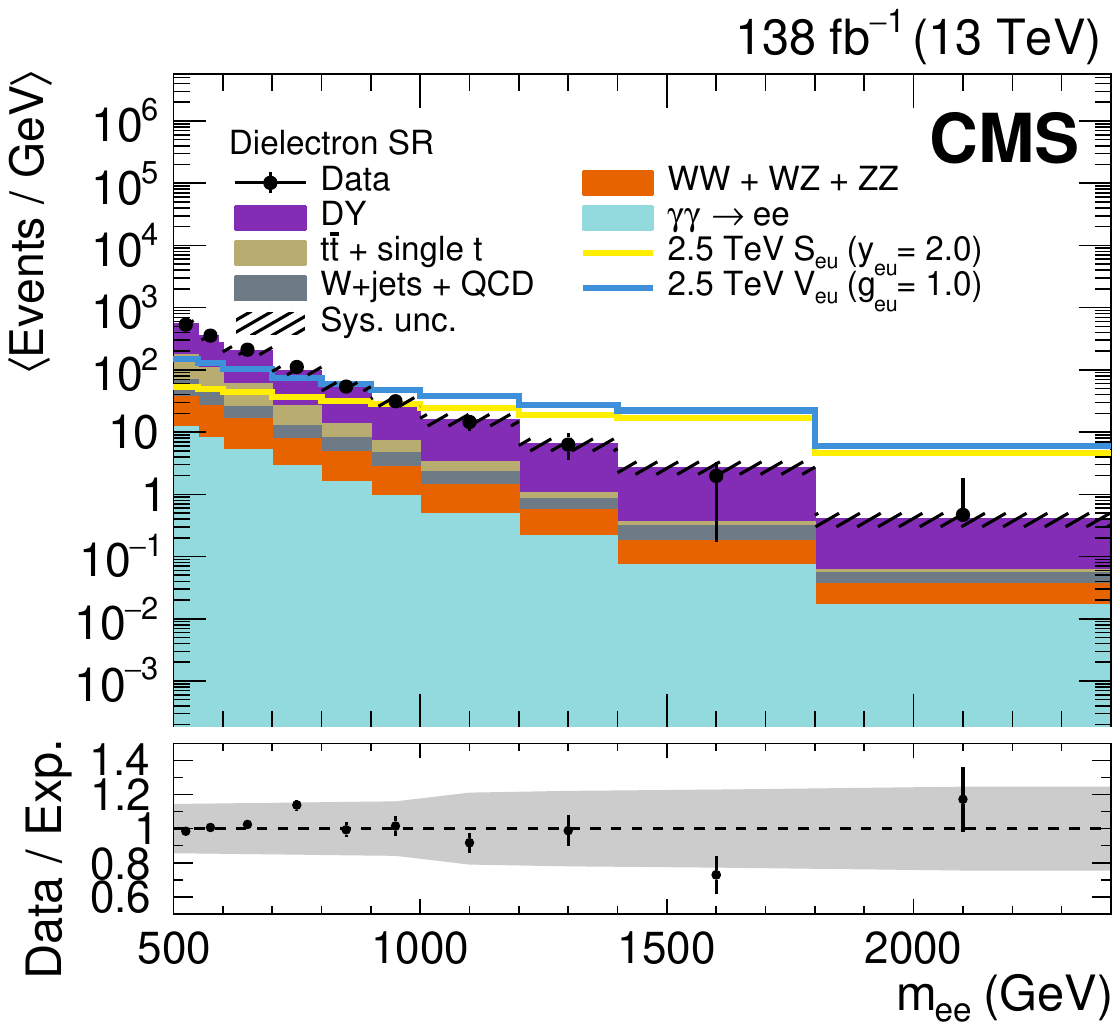} \\
  \includegraphics[width = 0.44 \textwidth]{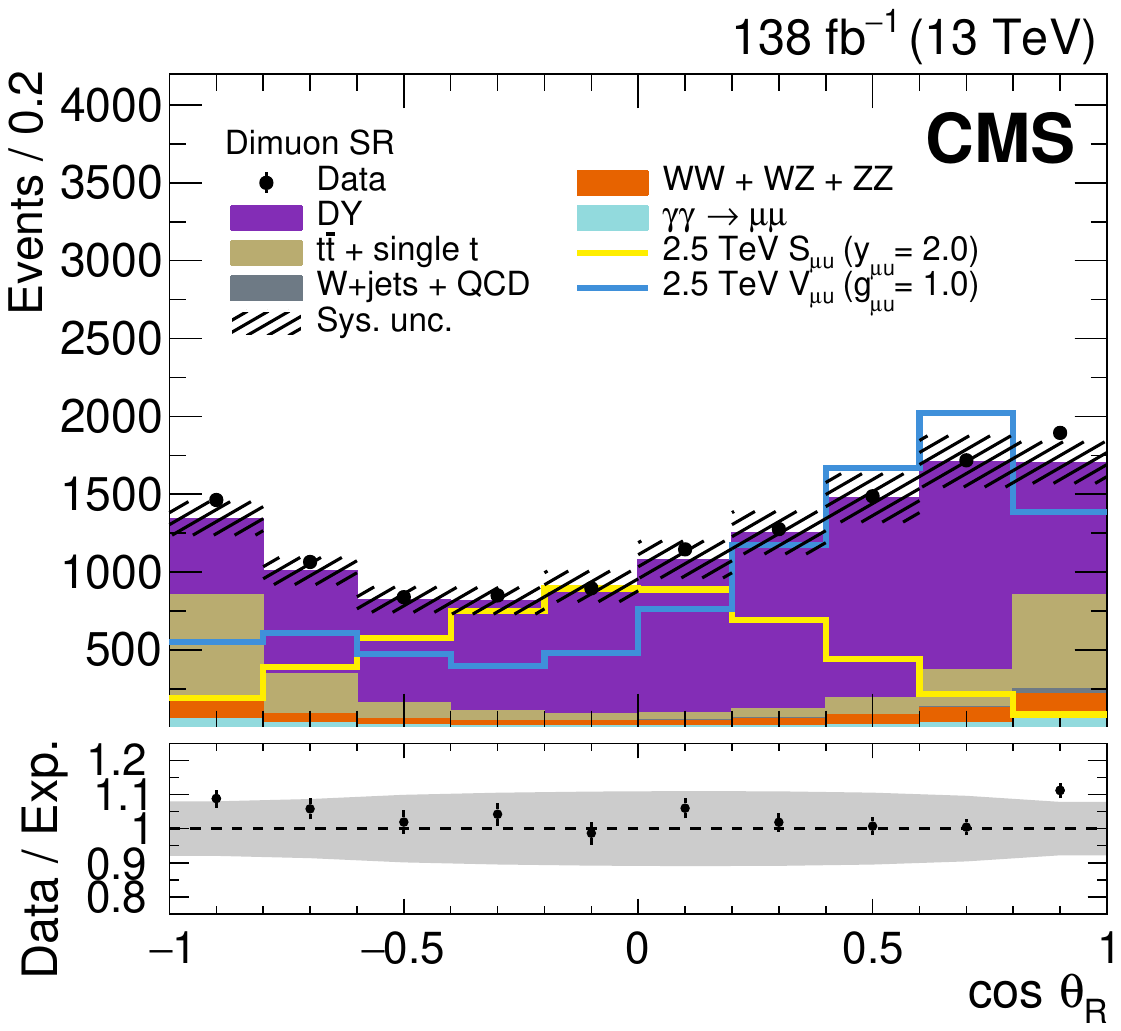}
  \includegraphics[width = 0.44 \textwidth]{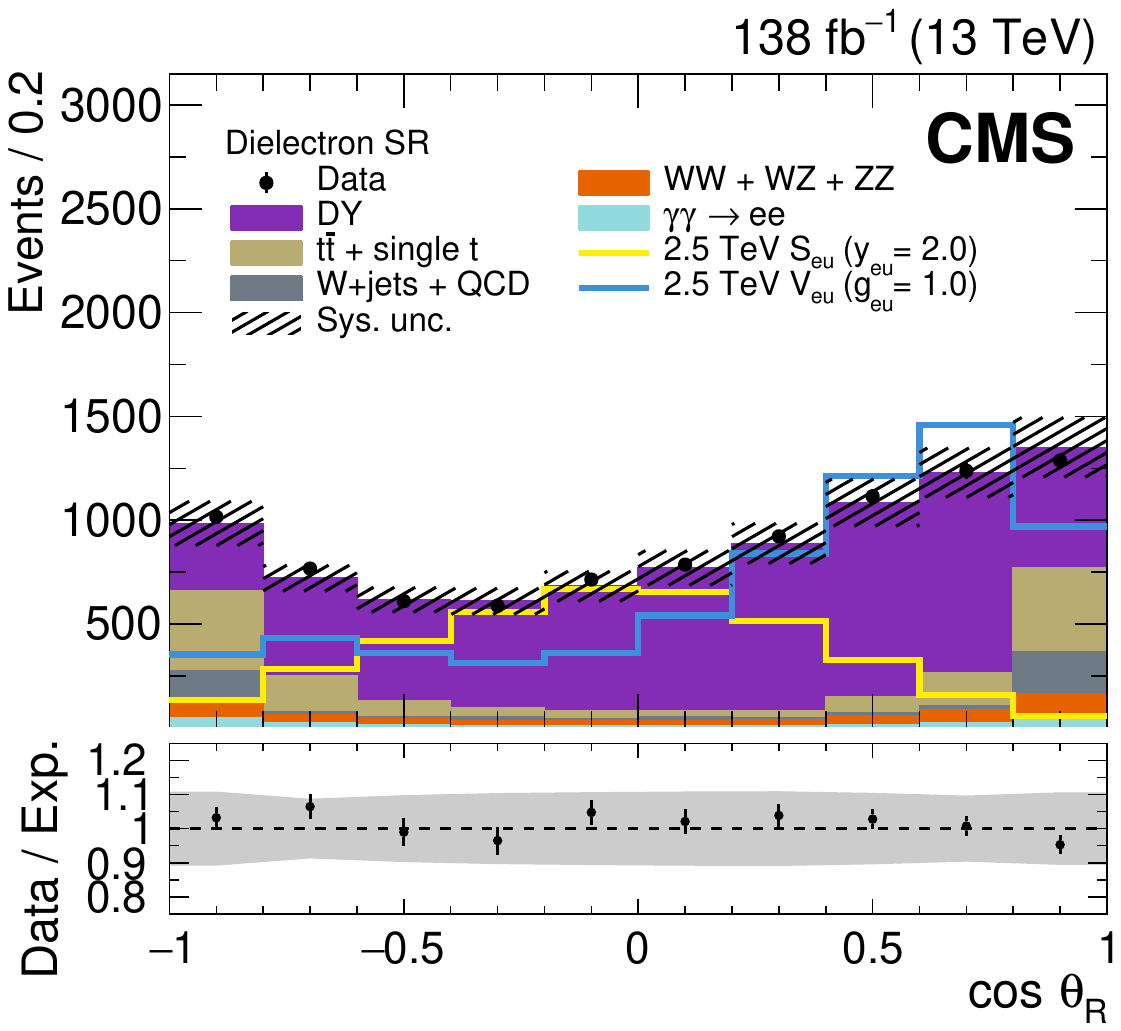}\\
  \includegraphics[width = 0.44 \textwidth]{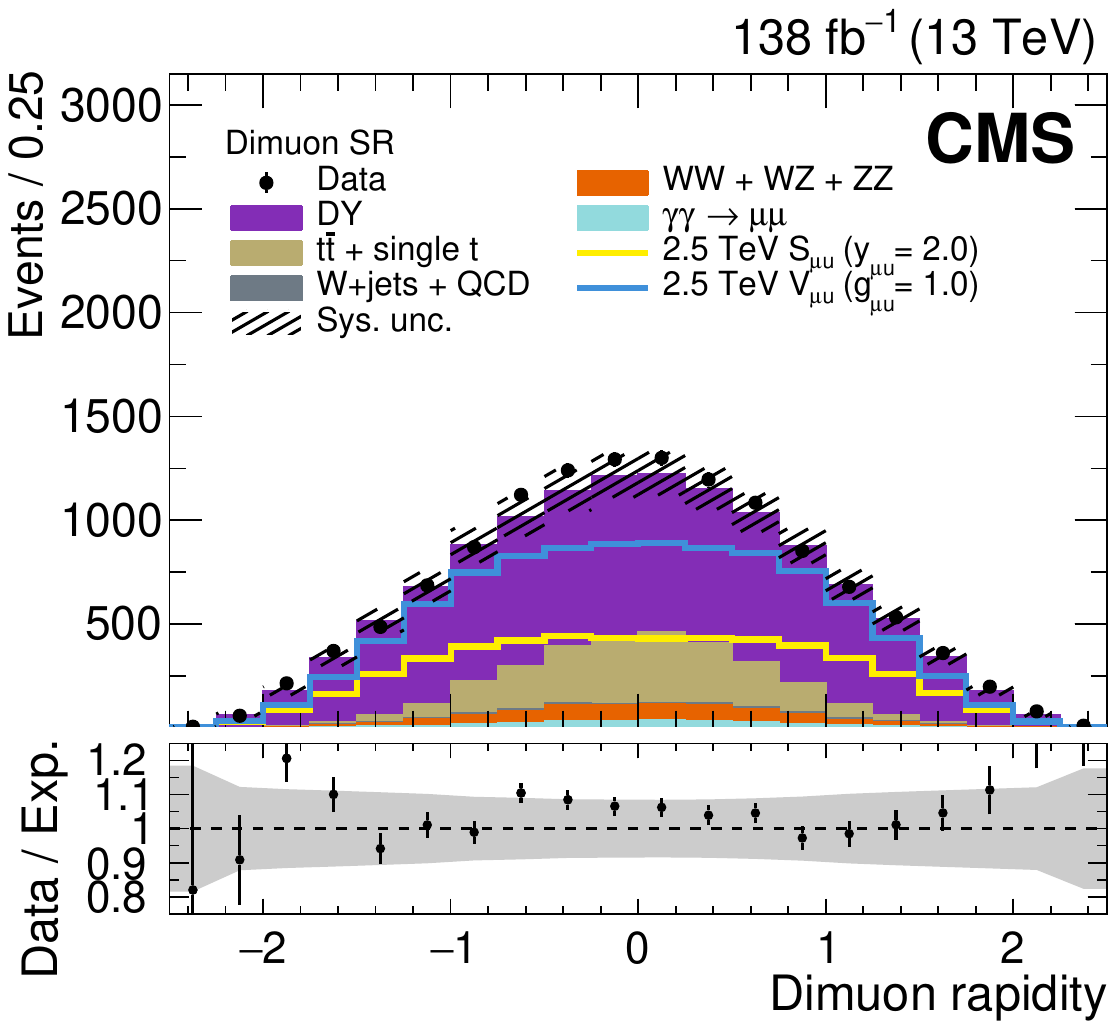}
  \includegraphics[width = 0.44 \textwidth]{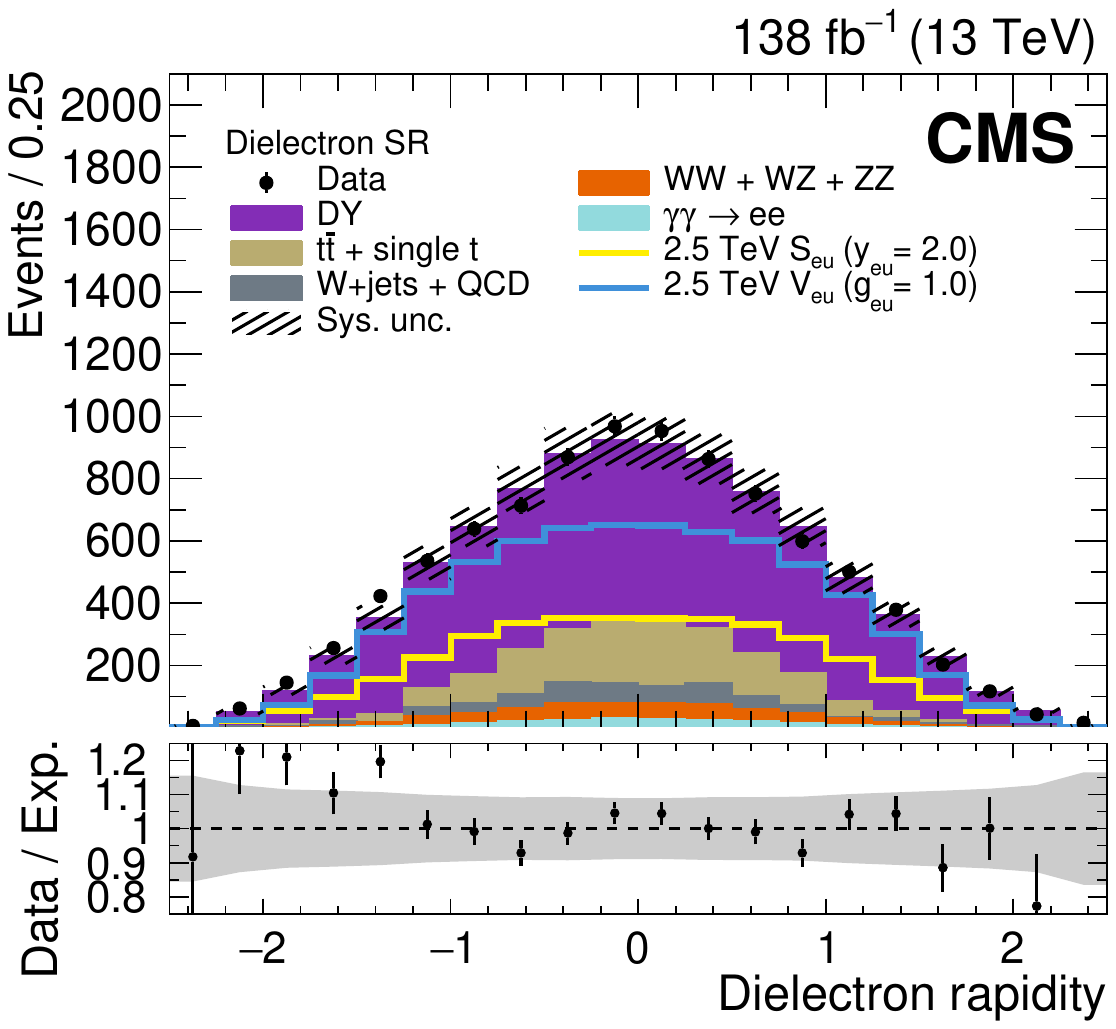}\\
  \caption{
    A comparison of data and expected background distributions in the dilepton invariant mass \mll (upper row), \csr (middle row), and dilepton rapidity (lower row). The distribution of \mll is shown as the average number of events per bin, whereas the distributions of \csr and rapidity are shown as the number of events per 0.2 in \csr and 0.25 in rapidity. 
    The left (right) plots show the \mumu (\ee) channel. 
    The blue histogram represents the signal yield of a hypothetical 2.5\TeV \PSmu (\PSeu) with $\ymu\,(\yeu)=2.0$, while the yellow histogram represents the signal yield of a hypothetical 2.5\TeV \PVmu (\PVeu) with $\gmu\,(\geu)=1.0$.
    The black points with error bars represent the data and their statistical uncertainties. 
    The background expectation is shown as stacked histograms.
    The hatched band shows the total systematic uncertainty in the expected background yield.
    The lower panels show the ratios of the data to the expectation. The gray bands represent the normalized uncertainty in the predicted yield. 
    The error bars in the ratio plot represent the normalized statistical uncertainty in the data.}
  \label{fig:Prefit_cmp}
\end{figure}

Figure~\ref{fig:Prefit_cmp} shows a comparison between measured \mumu and \ee events and the background prediction in the SR after all corrections have been applied.
Good agreement is observed between the simulated and observed numbers of \mumu and \ee events.

\section{Template construction} \label{sec:temp_construct}

To model the dilepton differential distributions, parameter-independent templates are constructed for each piece 
of the differential cross section of LQ production shown in Eqs.~\eqref{eq:ang_xsec_reparam_LQ} and \eqref{eq:ang_xsec_reparam_LQ_vec}.
For the SM DY component, the differential cross section is reparametrized as follows:
\begin{equation}
  \label{eq:ang_xsec_reparam}
  \left[\frac{\rd \sigma}{\rd \cstar}(\mll^2)\right]_\mathrm{DY} \propto \frac{3}{4(2+\alpha)}\biggl\lbrace (1 + c_{*}^2)  +  \frac{ (2 + \alpha)}{2} \;  \Afour \; c_{*} + \alpha (1 - c_{*}^2)\biggr\rbrace,
\end{equation}
where $\alpha = \frac{2 \Azero}{2 - \Azero}$. 
The first two terms are symmetric and antisymmetric in \cstar, respectively, and the third term is proportional to $\alpha$. 
The LQ exchange comprises two terms with coefficients $\ylq^4$ or $\glq^4$ for the pure exchange of the LQ and, $\ylq^2$ or $\glq^2$ for the interference with \Zg. 
The templates for these five pieces of the cross section are denoted as \fs (symmetric), \fa (antisymmetric), and \falpha (corresponding to the third term) for the three DY templates and \flqp and \flqi for the pure and interference LQ terms, respectively.

In Ref.~\cite{CMS:2022uul}, the $\fs$ , $\fa$ , and \falpha templates are constructed by binning events in \csr and reconstructed $\abs{y}$. These templates are constructed for various mass regions, and the fit is performed for each mass region separately.  
In this paper, the two-dimensional templates are extended to three dimensions by additionally binning in the reconstructed \mll. 
The templates are constructed by reweighting simulated DY events as analytical functions of generator-level quantities to match the differential distributions of the three pieces of the DY cross section. 
To reduce statistical fluctuations in the simulator, each event is used twice, once with $+\cstar$ and once with $-\cstar$, with a weight of 0.5 to keep the normalization unchanged. 
The reweighting functions for the DY templates, $\fs$ , $\fa$ , and \falpha respectively, are:
\begin{align}
  w_\mathrm{s}(\abs{\cstar}) &= \frac{1 + \cstar^2}{1+\cstar^2+\alpha\left(1-\cstar^2\right)}, \label{eq:sym_reweight} \\
  w_\mathrm{a}(\cstar) &= \frac{\cstar}{1+\cstar^2+\alpha\left(1-\cstar^2\right)}, \text{ and} \label{eq:asym_reweight} \\
  w_{\alpha}(\abs{\cstar}) &= \frac{1 - \cstar^2}{1+\cstar^2+\alpha\left(1-\cstar^2\right)}. \label{eq:alpha_reweight}
\end{align}
The denominator of the DY reweighting functions is determined for each event from distributions of simulated SM DY events that are binned in \costhetastar and generator-level $\abs{y}$.
The $\alpha$ values in the denominator are determined by fitting the generator-level distributions of the simulated events. 
Finally, to avoid negative values in the antisymmetric template, \fa, a linear combination of \fs and \fa is used in the fit, defined by $f_{\pm}=\frac{\fs \pm \fa}{2}$. 

For the templates corresponding to the two LQ terms, \flqp and \flqi, events are re\-weighted as functions of both \cstar and \mll. 
The reweighting functions are given by:
\begin{align}
  w_\mathrm{LQ(pure)}^{S,V}(\cstar,\mll) &= \left( \Npure^{S,V}(\mll)\frac{(1\mp \cstar)^2}{\left(1-\cstar+\frac{2\mlq^2}{\mll^2}\right)^2 }\right) \frac{1}{N_{\text{SM}}(\mll) (1+\cstar^2) } \label{eq:LQpure_reweight}
 \\
 \intertext{and}
  w_\mathrm{LQ(int)}^{S,V}(\cstar,\mll) &= \left( \Nint^{S,V}(\mll)\frac{(1\mp \cstar)^2}{\left(1-\cstar+\frac{2\mlq^2}{\mll^2}\right)} \right) \frac{1}{N_{\text{SM}}(\mll) (1+\cstar^2)}, \label{eq:LQint_reweight}
\end{align}
where $S$ and $V$ denote the scalar or vector case. 
The prefactors, $N_{\text{pure/int}}^{S,V}(\mll)$, depend on the vector and axial-vector couplings of the quarks and leptons to the \PZ boson, as well as the quark charges. 
Thus, templates for an LQ coupled to a lepton and quark of specific flavors have different shapes compared to an LQ coupled to a different lepton and quark. 
The $N_{\text{SM}}(\mll)$ prefactor represents the coefficient of the symmetric term $(1+\cstar^2)$ in the SM DY angular distribution. 
The denominators of the LQ reweighting functions are different from those of the SM DY functions, calculated from the analytical form of the LO SM DY cross section for each event. 
This is because the LQ modifies the \mll distribution, which must be taken into account in the reweighting. 

The reweighting procedure described above has several benefits.
The effects of misassigning the direction of the incident quark in the $\cstar$ computation, called the ``dilution'' effect, are accounted for correctly. 
Additionally, no dedicated MC samples need to be produced for the LQ process or its interference. 
The distribution of LQ events can be obtained by reweighting SM DY MC events using the analytical functional form of the LQ differential cross section. 

The reweighting procedure was further validated by fitting LQ and DY templates at the generator level to simulated LQ signals using \MGvATNLO using the \textsc{SLQrules} package~\cite{Crivellin:2021ejk}. Signals were generated at various mass points from 1--5\TeV. The fitted results agreed with the expected coupling values up to a maximum of 30\% at $\mlq=1\TeV$, decreasing with \mlq. The discrepancy is attributable in part to the fact that the reweighting procedure does not account for NLO effects in the LQ exchange. We account for this in our fits by applying a 30\% systematic uncertainty in the normalization of the LQ templates.

The total template for the scalar LQ case, binned in reconstructed \mll, $y$, and \csr, is given by:
\begin{equation}
\begin{aligned}
  f_{\text{data}} =& \sum_j f^j_\text{bkg} + N(\alpha) \left( \alpha \; \falpha +  \left(1 + \frac{3\Afour}{8N}\right)f_{+} + \left(1 - \frac{3\Afour}{8N}\right)f_{-}\right) \\
  &+ \ylq^4\flqp + \ylq^2\flqi,
  \label{eq:template_fit}
\end{aligned}
\end{equation}
where $f^j_{\text{bkg}}$ are templates for the non-DY backgrounds and $N(\alpha) = \frac{3}{4(2+\alpha)}$. 
The same template is used for the vector LQ case using the corresponding reweighting functions, and replacing \ylq with \glq. 
The coefficients \Afour, $\alpha$, and either $\ylq^2$ or $\glq^2$ are extracted in the fits to data. 

Events are divided into bins based on \mll, $\abs{y}$, and \csr. 
For \mll, three bins are used with bin edges at 500, 700, and 1000\GeV, with the third bin containing all events with $\mll>1000\GeV$.
Three bins are defined for $\abs{y}$, with edges at 0, 0.6, 1, and 2.4. 
The binning in \csr depends on the rapidity.
Within the first rapidity bin, events are divided into eight bins in \csr of width 0.25. 
For the other two rapidity bins, events are divided into six bins in \csr, with edges of $-1$, $-0.5$, $-0.25$, $0$, $0.25$, $0.5$, and $1$. Bin edges were chosen after ensuring that there were sufficient simulated events in each bin, and fit uncertainties were minimized without losing precision.

Additional templates are created for the other background processes, with one template each for the MisID background, photon-induced dilepton production, DY \tautau production, the combined top quark backgrounds (\ttbar, \tW), and the combined diboson backgrounds ($\PW\PW, \PZ\PZ, \PW\PZ$). 
The top quark and MisID templates are symmetrized in \csr to reduce statistical fluctuations. 
The diboson, photon-induced dilepton,  and DY \tautau backgrounds are not symmetrized because of their significant inherent asymmetry. 

\section{Systematic uncertainties}\label{sec:sys_uncs}

Systematic uncertainties in the normalization and shapes of the templates arise from a variety of sources, 
and are incorporated in the fit to data through nuisance parameters in the likelihood. 
Systematic uncertainties affecting template normalizations are included using log-normal priors.
For systematic uncertainties that change the shape of a template, shifted templates are constructed by varying 
the source of the systematic uncertainty up and down within its uncertainty. 
The uncertainties are incorporated into the likelihood by interpolation between the nominal and shifted templates, constrained with a Gaussian prior. 
Table~\ref{tab:sep_sys_uncs_2} shows the impact of each uncertainty in the fitted LQ coupling for representative scalar and vector signal hypotheses. 

\begin{table}[!htbp]
  \centering
  \topcaption{
    The contributions of the statistical uncertainty and of individual sources of systematic uncertainty to the total variance in the fitted value of $\ylq^2$ ($\glq^2$) for a scalar (vector) \mlq of 2.5\TeV. 
    For a given source of uncertainty, the impact is determined by fixing its associated nuisance parameter to the postfit value and evaluating the change in the total uncertainty. The quadrature sum of the uncertainties is approximately equal to the total uncertainty in $\ylq^2$ ($\glq^2$) for a scalar (vector) \mlq of 2.5\TeV.
  }
    \label{tab:sep_sys_uncs_2}
    \renewcommand{\arraystretch}{1.2}
    \cmsTable{
      \begin{tabular}{l c c c c c c c c}
        Source   & \multicolumn{8}{c}{Relative contribution (\%)} \\
        & $\yeu^2$ & $\yed^2$ & $\ymu^2$ & $\ymd^2$ & $\geu^2$ & $\ged^2$ & $\gmu^2$ & $\gmd^2$ \\
        \hline
          Statistical uncertainty & 84 & 84 & 81 & 79 & 34 & 33 & 25 & 32 \\ 
          LQ LO reweighting & 21 & 31 & 27 & 25 & 28 & 27 & 53 & 37 \\ 
          MC and MisID backgrounds statistical uncertainty & 15 & 15 & 14 & 16 & 35 & 56 & 31 & 46 \\ 
          \alpS + renormalization/factorization scales & 7 & 11 & 20 & 22 & 15 & 19 & 28 & 32 \\ 
          \emu shape corrections & 7 & 8 & 17 & 16 & 11 & 4 & 16 & 11 \\ 
          Background cross sections & 12 & 14 & 33 & 36 & 23 & 22 & 54 & 48 \\ 
          Integrated luminosity & 13 & 13 & 20 & 23 & 19 & 19 & 32 & 38 \\ 
          Pileup & 6.9 & 6.4 & 9.5 & 8.8 & 5.1 & 5.7 & 5.1 & 6.0 \\ 
          PDFs & 6.0 & 6.9 & 7.1 & 8.2 & 8.8 & 4.6 & 11 & 8.8 \\ 
          Trigger efficiency & 11 & 9.0 & 2.5 & 1.3 & 19 & 7.1 & 4.3 & 6.3 \\ 
          Lepton momentum scale & 17 & 22 & 5.7 & 3.1 & 30 & 12 & 14 & 6.9 \\ 
          Lepton identification/isolation & 33 & 20 & 9.9 & 10 & 62 & 57 & 12 & 16 \\ 
          MisID shape \& normalization  & 17 & 19 & 15 & 16 & 38 & 58 & 31 & 47 \\
    \end{tabular}
  }
\end{table}

A significant source of uncertainty is due to the creation of the LQ signal templates using LO reweighting of NLO DY MC events. 
A normalization uncertainty of 30\% in the coupling is assigned to the pure and interference LQ templates. This uncertainty is based on bias tests performed on simulations with injected signal events that were generated with \MGvATNLO following Ref.~\cite{Crivellin:2021ejk}. This uncertainty does not affect the shape of the templates.

Several sources of systematic uncertainty are due to the modeling of leptons in simulation. 
These include uncertainties in the momentum scale, the efficiency of the reconstruction and identification criteria, and the trigger efficiency. 
These uncertainties are small for the muon channel, but are dominant in the electron channel. 

Additional systematic uncertainties are applied to templates estimated from simulation. 
The uncertainty in the PDFs is assigned from the Hessian variations of the PDF sets. 
Uncertainties due to missing higher-order contributions are estimated by varying the renormalization and factorization scales independently by factors of two. 
Additional uncertainties are assigned to the cross sections of the DY, \ttbar, and diboson processes based on comparisons between generators, of size 3, 5, and 4\%, respectively. 
Similarly, an uncertainty of 6\% is assigned to the normalization of the $\PGg\PGg\to\Pell\Pell$ process, based on a comparison between the Suri--Yennie structure functions and the LuxQED photon PDFs~\cite{Manohar:2016nzj,Manohar:2017eqh}. 
The statistical uncertainties due to the limited number of MC events are implemented using the simplified Barlow--Beeston approach~\cite{Conway:2011in}. Since the bin contents of $\pm \cstar$ are correlated because each event is used twice, each pair of $\pm \cstar$ bins is varied together and accounts for a separate nuisance parameter.

The MisID background is assigned a normalization uncertainty of 50\% based on closure studies of the misidentification rate in simulation and a shape uncertainty based on the difference between the MisID estimate in SS and OS events. 
The MisID uncertainty primarily affects the electron channel, given the higher rate of misidentified electrons than muons.

The \ttbar and diboson templates are assigned a shape uncertainty due to the correction to the \csr distributions derived from \emu events. 
The integrated luminosities for the 2016, 2017, and 2018 data-taking years have 1.2--2.5\% individual uncertainties~\cite{CMS-LUM-17-003,CMS-PAS-LUM-17-004,CMS-PAS-LUM-18-002}, while the overall uncertainty for the 2016--2018 period is 1.6\%.
Small uncertainties in the modeling of pileup are also incorporated.  

The uncertainties in the PDFs, pileup, electron and muon efficiencies, MC cross sections, luminosity, and the LQ template normalization are correlated for all the three data-taking years. The \emu shape corrections, and renormalization and factorization scale corrections are correlated for the 2017 and 2018 periods. The rest are treated independently per year. In the results for the scalar LQ couplings, the contribution of statistical uncertainties is larger than the systematic uncertainties. However for the vector LQ couplings, the systematic uncertainties have a larger contribution. This is because of the differences in shapes between the scalar and vector LQ templates. 

Table~\ref{tab:sep_sys_uncs_2} shows that the largest contributors to the uncertainties in $\ymu^2$, $\ymd^2$, $\gmu^2$, and $\gmd^2$ are those arising from the LO reweighting scheme, the renormalization and factorization scales, and the background cross sections. Additionally, the electron identification and isolation has major contributions to the uncertainties in $\yeu^2$, $\yed^2$, $\geu^2$, and $\ged^2$. The post-fit up and down variations of these systematic uncertainties are asymmetric in their impacts on the LQ couplings. This is because the likelihood function for each model is asymmetric in $\ylq^2\,(\glq^2)$, and not quadratic. The shapes of the likelihoods follow from Eqs.~\eqref{eq:ang_xsec_reparam_LQ} and \eqref{eq:ang_xsec_reparam_LQ_vec}. For each LQ model considered, the prefactor $\Nint^{S(V)}$ can assume positive or negative values based on the fermions involved in the LQ exchange, leading to constructive or destructive interference. Accordingly, combinations of the pure LQ exchange term and the LQ-SM interference term result in asymmetric likelihoods, that become highly nonquadratic for large couplings.

\section{Results}\label{sec:results}

\begin{figure}[hbtp]
  \centering
    \includegraphics[width=0.49 \textwidth]{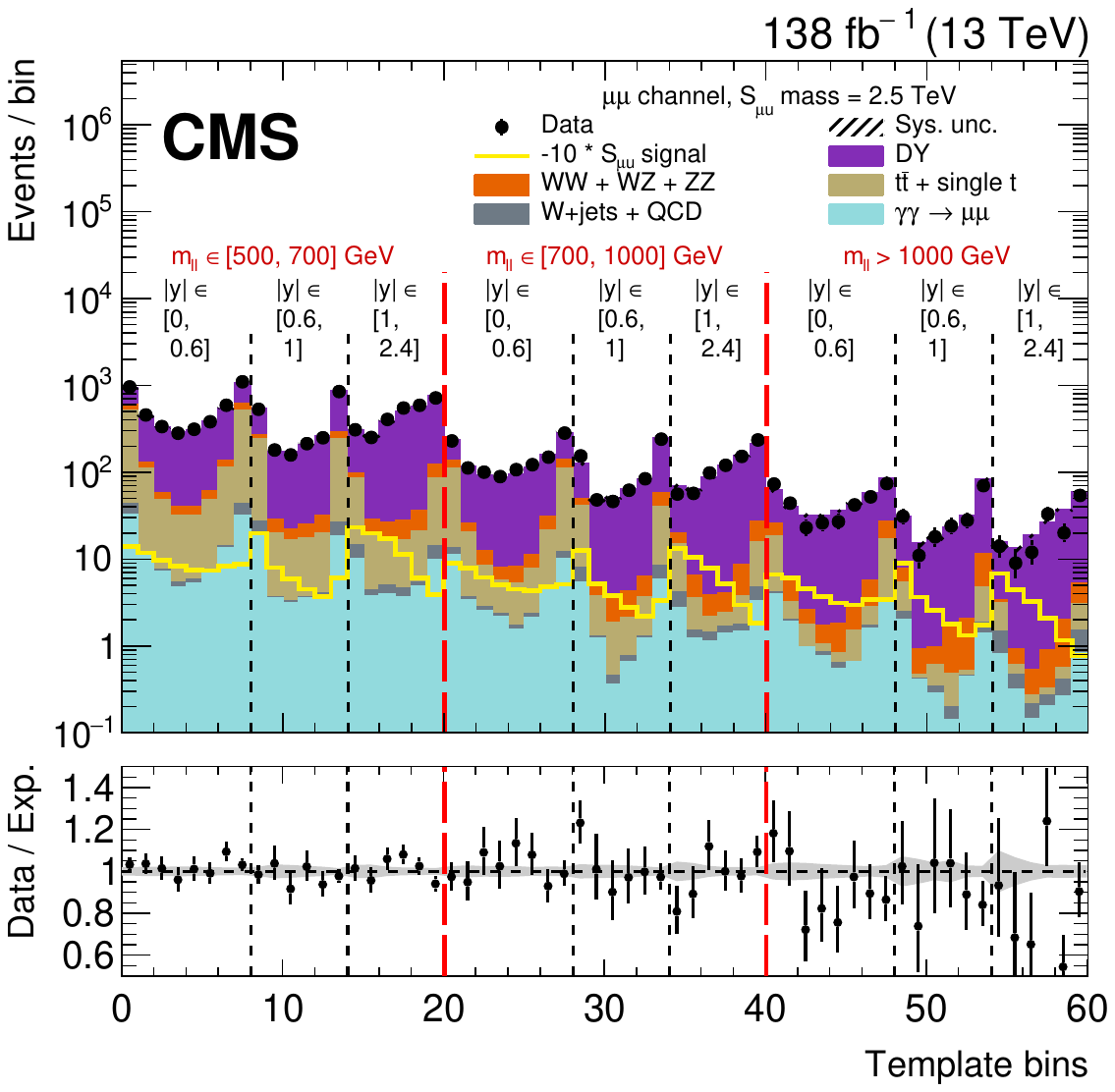}
    \includegraphics[width=0.49 \textwidth]{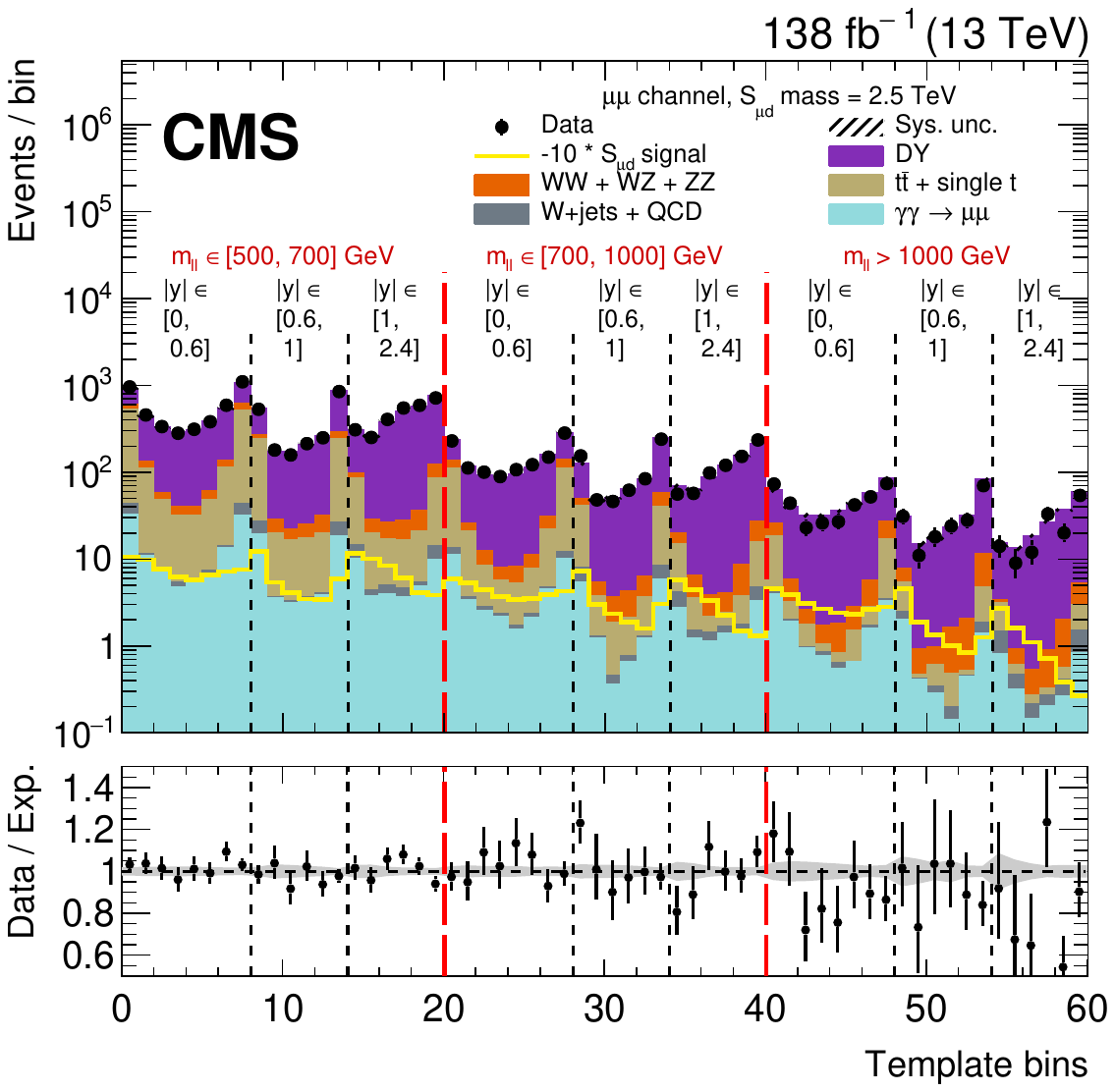}
    \caption{
    The observed data in the dimuon channel and the fitted signal-plus-background templates, shown for the \PSmu (left) and \PSmd (right) scenarios with a candidate \mlq of 2.5\TeV. 
    The black points are the observed data, the stacked histograms represent the backgrounds, and the yellow histogram shows the fitted LQ signal scaled by $-10$. Distributions of events are binned in the reconstructed \mll (vertical red dashed lines), $\abs{y}$ (vertical black dashed lines), and \csr (Section~\ref{sec:temp_construct}). The lower panels show the ratios of the data to the expectation. The gray
    bands represent the normalized uncertainty in the predicted yield. The error bars in the ratio plot represent the normalized statistical uncertainty in the data.
    }
\label{fig:unblind_scalar_results2}
\end{figure}

\begin{figure}[hbtp]
  \includegraphics[width = 0.49 \textwidth]{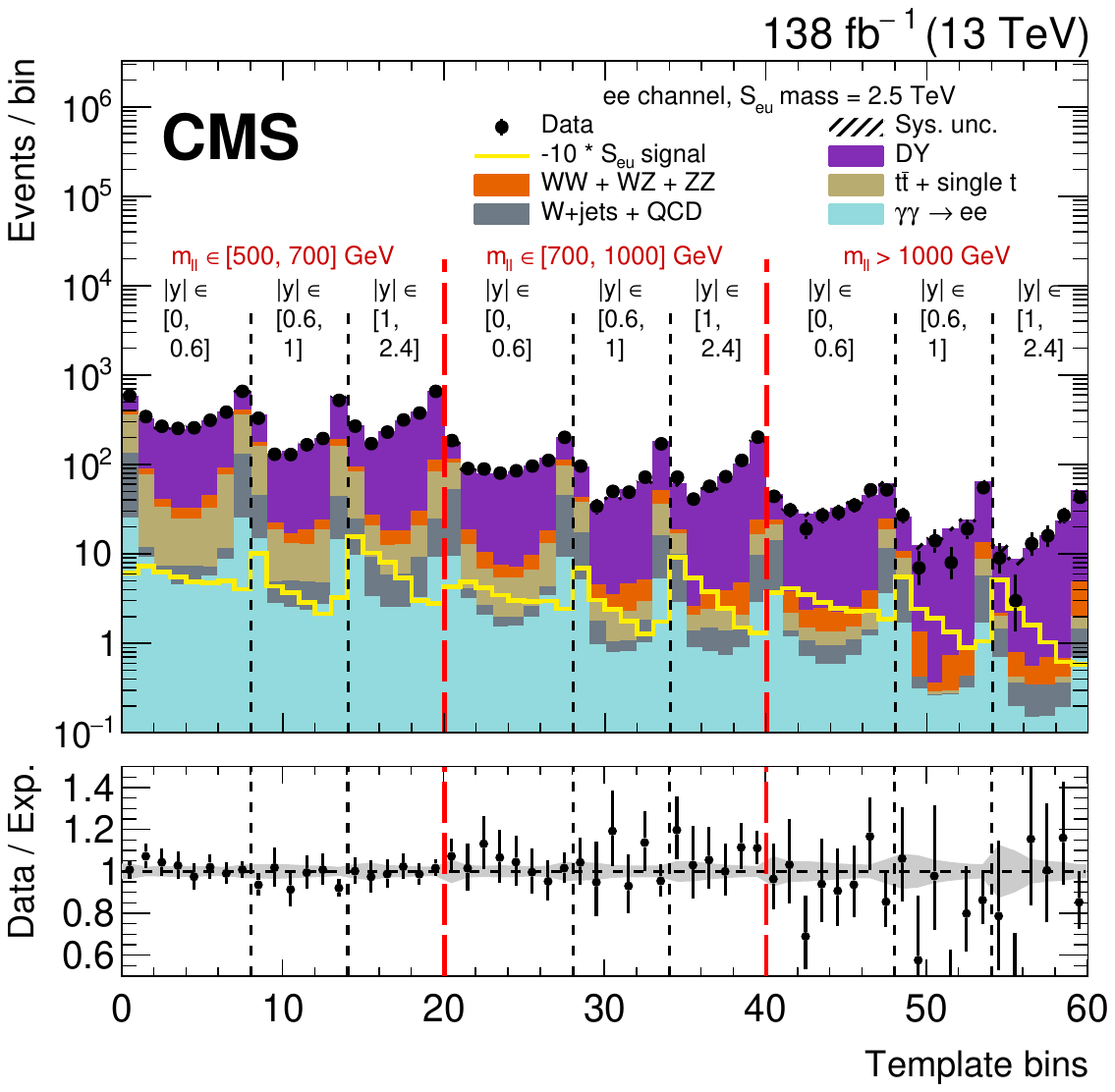}
  \includegraphics[width = 0.49 \textwidth]{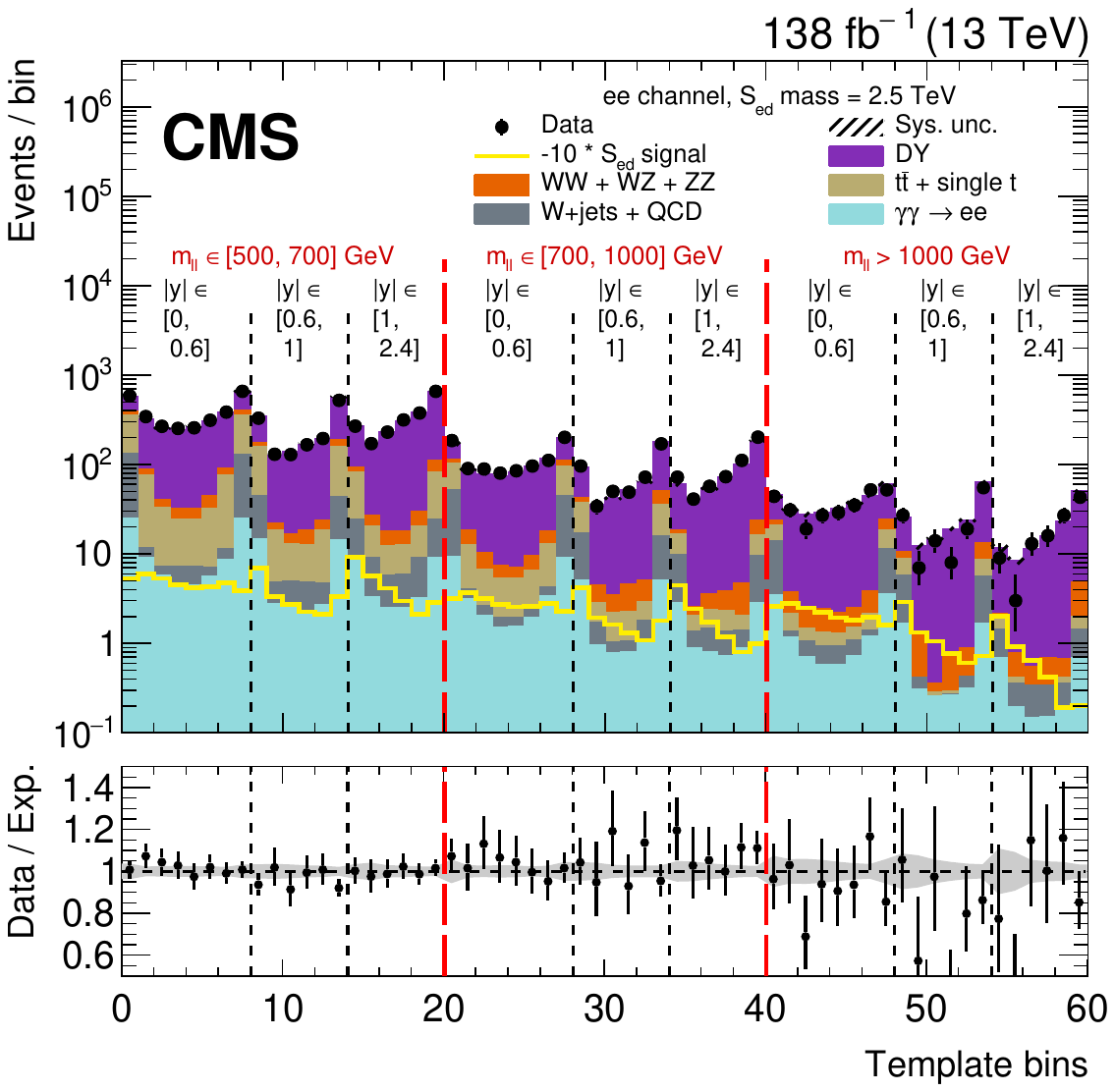}
   \centering
  \caption{
    The observed data in the dielectron channel and the fitted signal-plus-background templates, shown for the \PSeu (left) and \PSed (right) scenarios with a candidate \mlq of 2.5\TeV. 
    The black points are the observed data, the stacked histograms represent the backgrounds, and the yellow histogram shows the fitted LQ signal scaled by $-10$. Distributions of events are binned in the reconstructed \mll (vertical red dashed lines), $\abs{y}$ (vertical black dashed lines), and \csr (Section~\ref{sec:temp_construct}). The lower panels show the ratios of the data to the expectation. The gray
    bands represent the normalized uncertainty in the predicted yield. The error bars in the ratio plot represent the normalized statistical uncertainty in the data.
    }      
  \label{fig:unblind_scalar_results1}
\end{figure}

\begin{figure}[hbtp]
  \centering
  \includegraphics[width = 0.49 \textwidth]{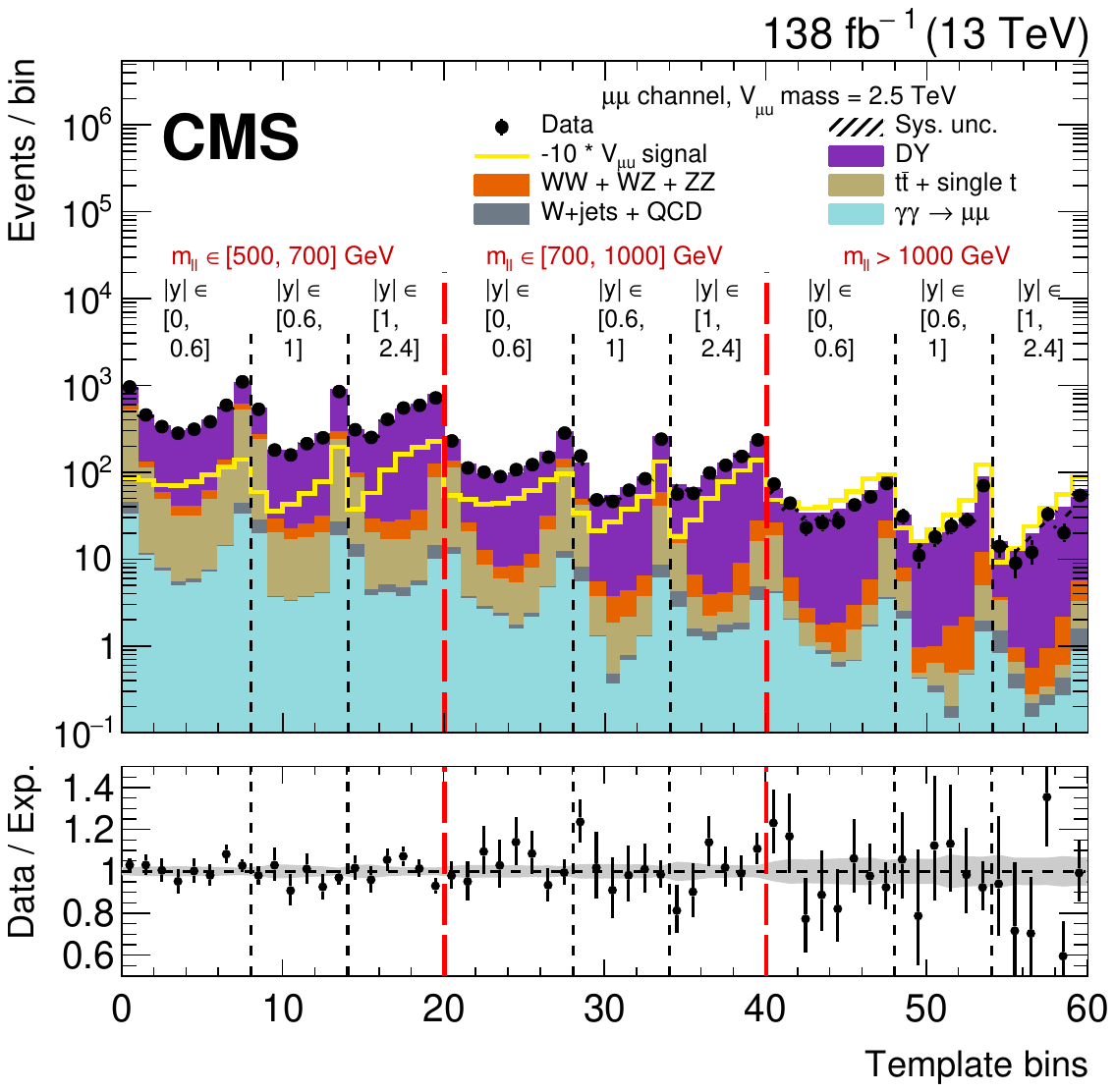}
  \includegraphics[width = 0.49 \textwidth]{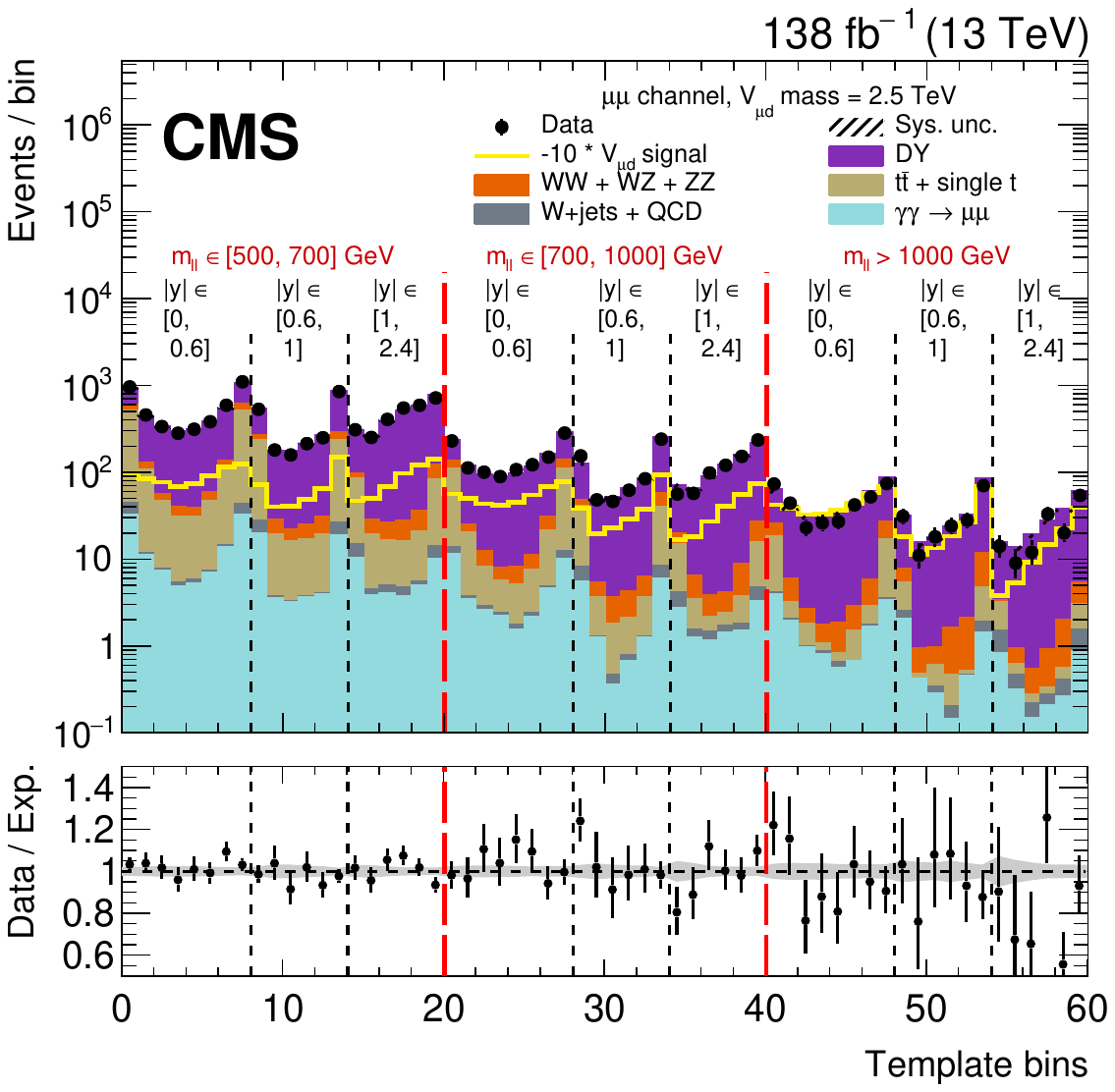}
  \caption{
    The observed data in the dimuon channel and the fitted signal-plus-background templates, shown for the \PVmu (left) and \PVmd (right) scenario with a candidate \mlq of 2.5\TeV. 
    The black points are the observed data, the stacked histograms represent the backgrounds, and the yellow histogram shows the fitted LQ signal scaled by $-10$. Distributions of events are binned in the reconstructed \mll (vertical red dashed lines), $\abs{y}$ (vertical black dashed lines), and \csr (Section~\ref{sec:temp_construct}). The lower panels show the ratios of the data to the expectation. The gray
    bands represent the normalized uncertainty in the predicted yield. The error bars in the ratio plot represent the normalized statistical uncertainty in the data.
    }
  \label{fig:unblind_vector_results2}
\end{figure}

\begin{figure}[hbtp]
  \centering
  \includegraphics[width = 0.49 \textwidth]{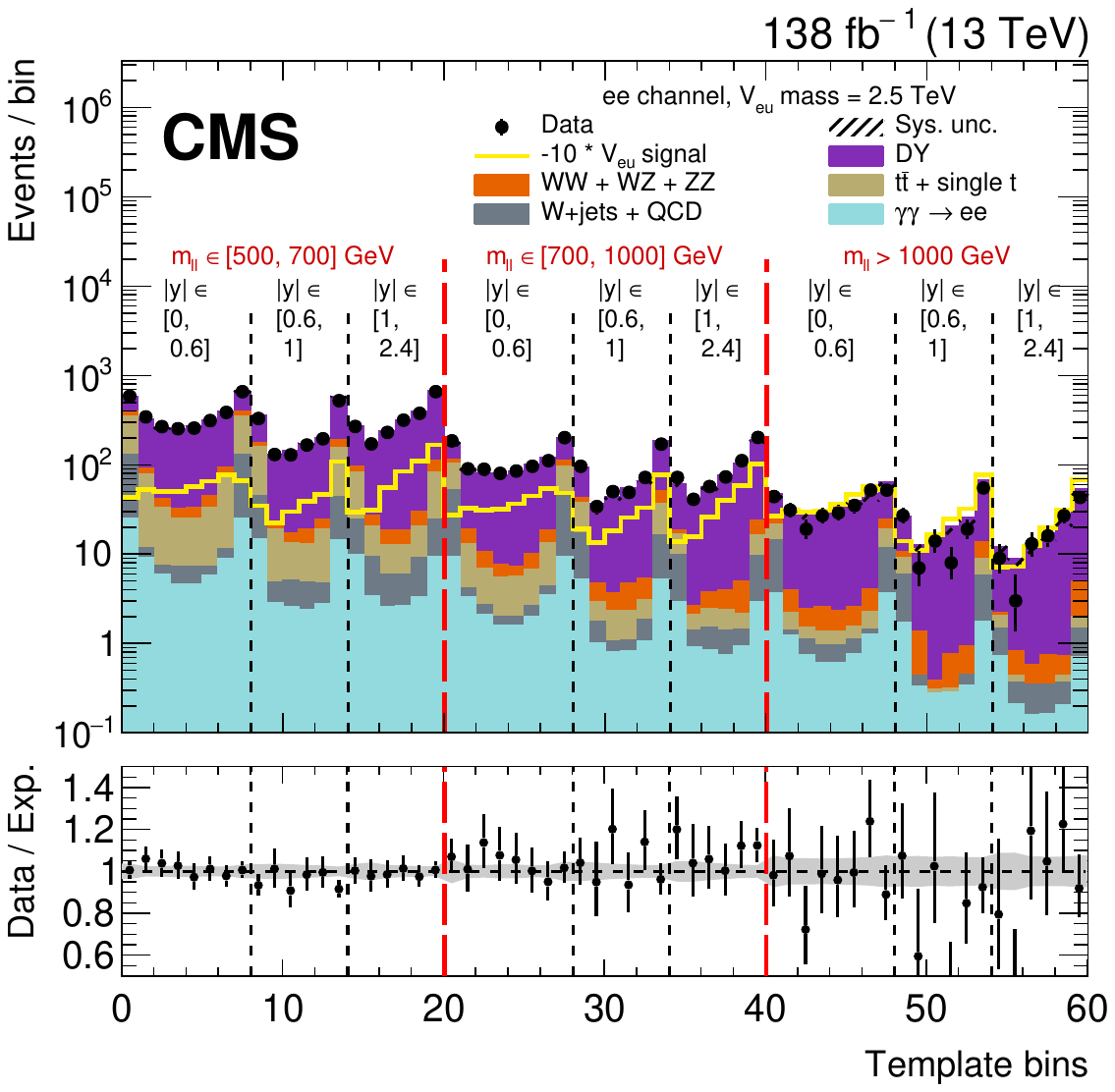}
  \includegraphics[width = 0.49 \textwidth]{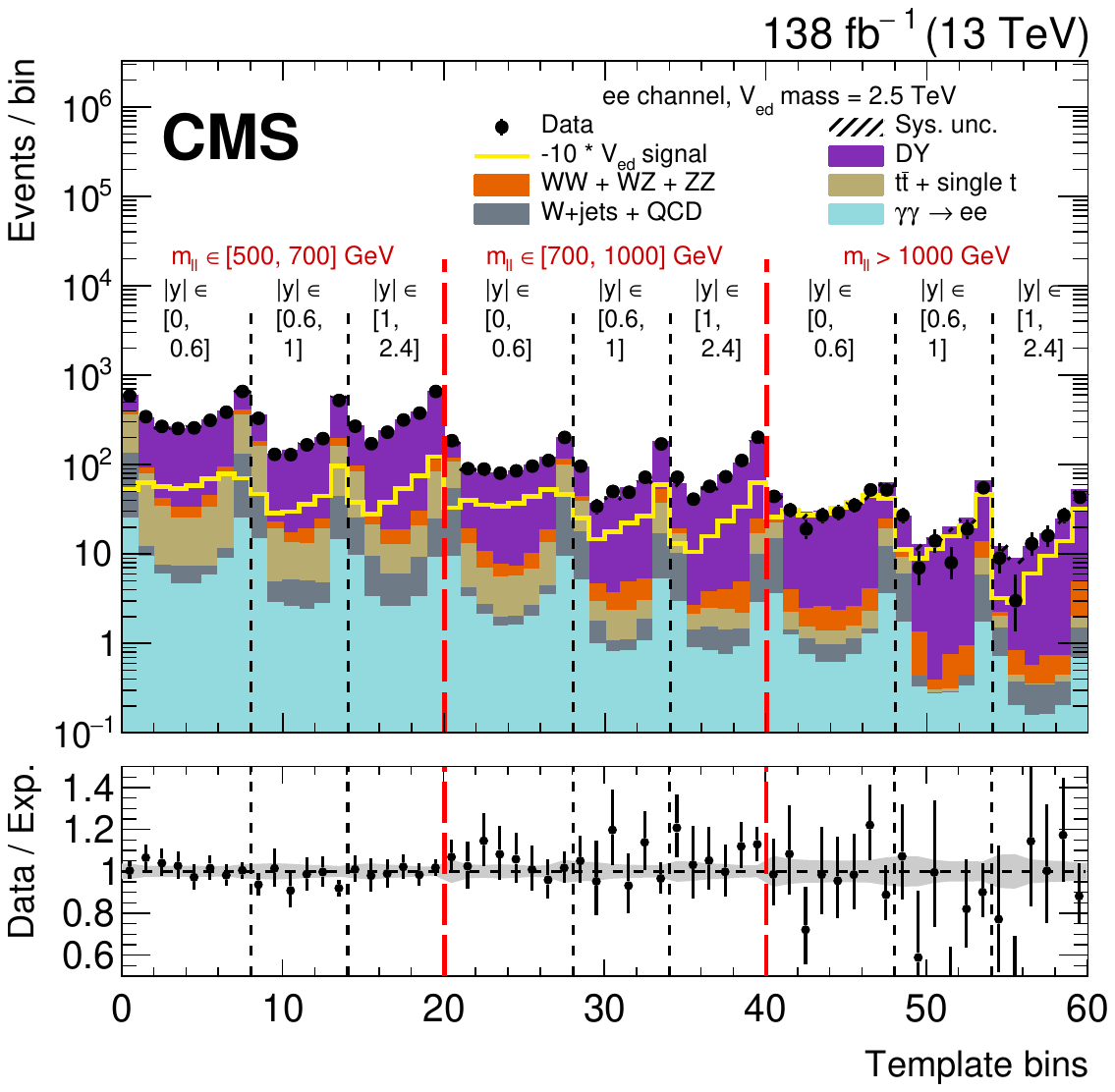}
  \caption{
    The observed data in the dielectron channel and the fitted signal-plus-background templates, shown for the \PVeu (left) and \PVed (right) scenario with a candidate \mlq of 2.5\TeV. 
    The black points are the observed data, the stacked histograms represent the backgrounds, and the yellow histogram shows the fitted LQ signal scaled by $-10$. Distributions of events are binned in the reconstructed \mll (vertical red dashed lines), $\abs{y}$ (vertical black dashed lines), and $\csr$ (Section~\ref{sec:temp_construct}). The lower panels show the ratios of the data to the expectation. The gray
    bands represent the normalized uncertainty in the predicted yield. The error bars in the ratio plot represent the normalized statistical uncertainty in the data.
    }
  \label{fig:unblind_vector_results1}
\end{figure}

Template fits are performed to the data for each LQ signal hypothesis, following Eq.~\eqref{eq:template_fit}. The \textsc{Combine} package~\cite{CMS:2024onh} is used to perform maximum likelihood fits to the data.
 A range of positive and negative values of $\ylq^2$ $(\glq^2)$ is floated in the fit in order to ensure fit stability. The extracted best fit parameters for a candidate \mlq of 2.5\TeV are shown in Tables~\ref{tab:results_scalar} and \ref{tab:results_vector}. The corresponding upper and lower bounds of the Feldman--Cousins intervals~\cite{FC} for $\ylq^2$ $(\glq^2)$ are reported as well. In the SM, the expected value of \Azero is 0.05, and that of \Afour is 1.61. These quantities were measured more precisely in Ref.~\cite{CMS:2022uul}. The best fit values for $\ged^2$ and $\gmd^2$ are positive, while the others are negative. This is due to the destructive interference between LQ and SM amplitudes for $\PVed$ and $\PVmd$ exchanges, and constructive interference in the other models. All results are found to be consistent with the SM background expectations. The fit results are shown in Figs.~\ref{fig:unblind_scalar_results2}--\ref{fig:unblind_vector_results1}. In each figure, the overall signal yield is scaled by $-10$ because the sum of the pure and interference contributions scaled by the post-fit values of $\ylq^4$ $(\glq^4)$ and $\ylq^2$ $(\glq^2)$ is negative. 

\begin{figure}[hbtp]
  \includegraphics[width = 0.49 \textwidth]{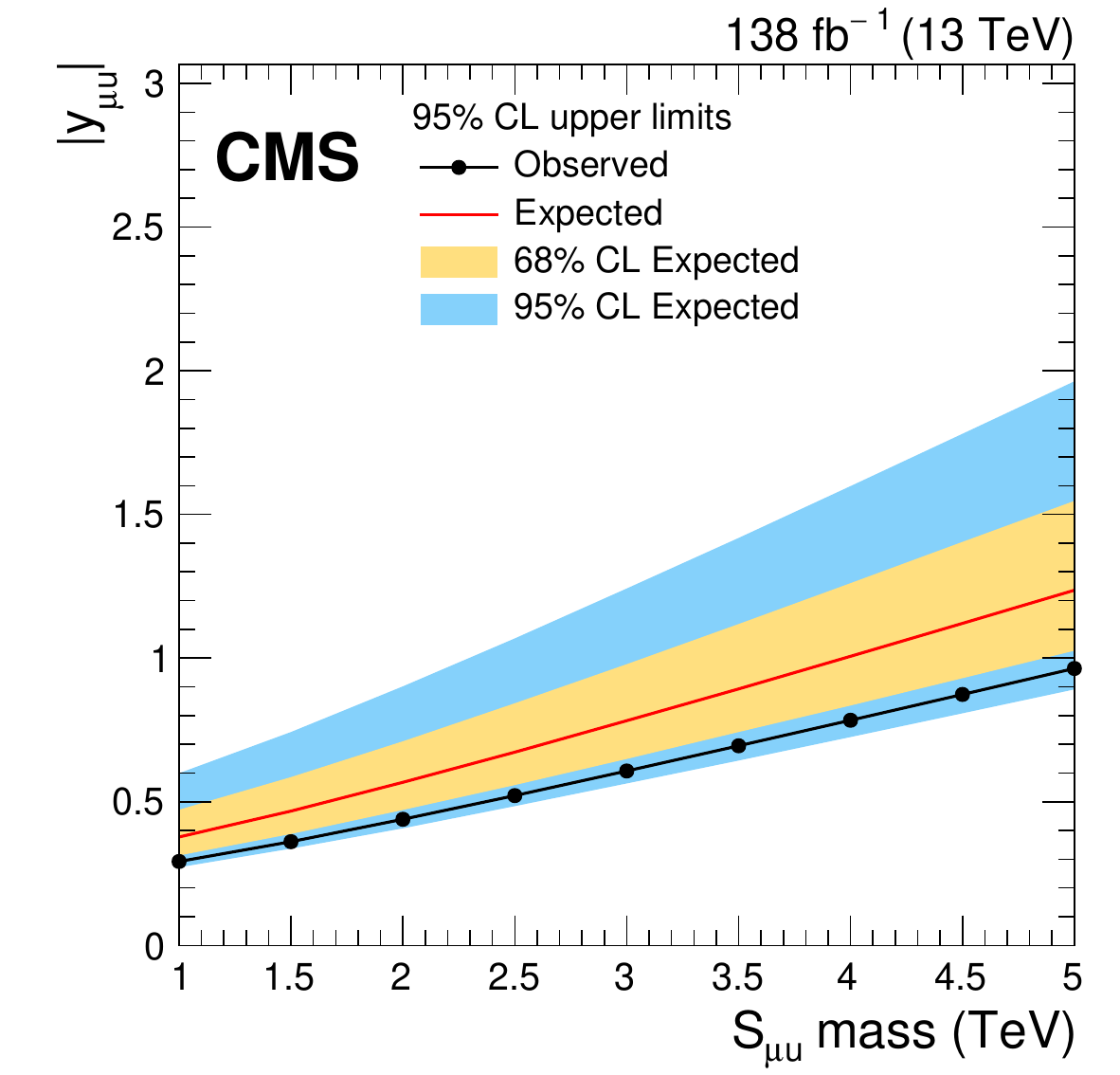}
  \includegraphics[width = 0.49 \textwidth]{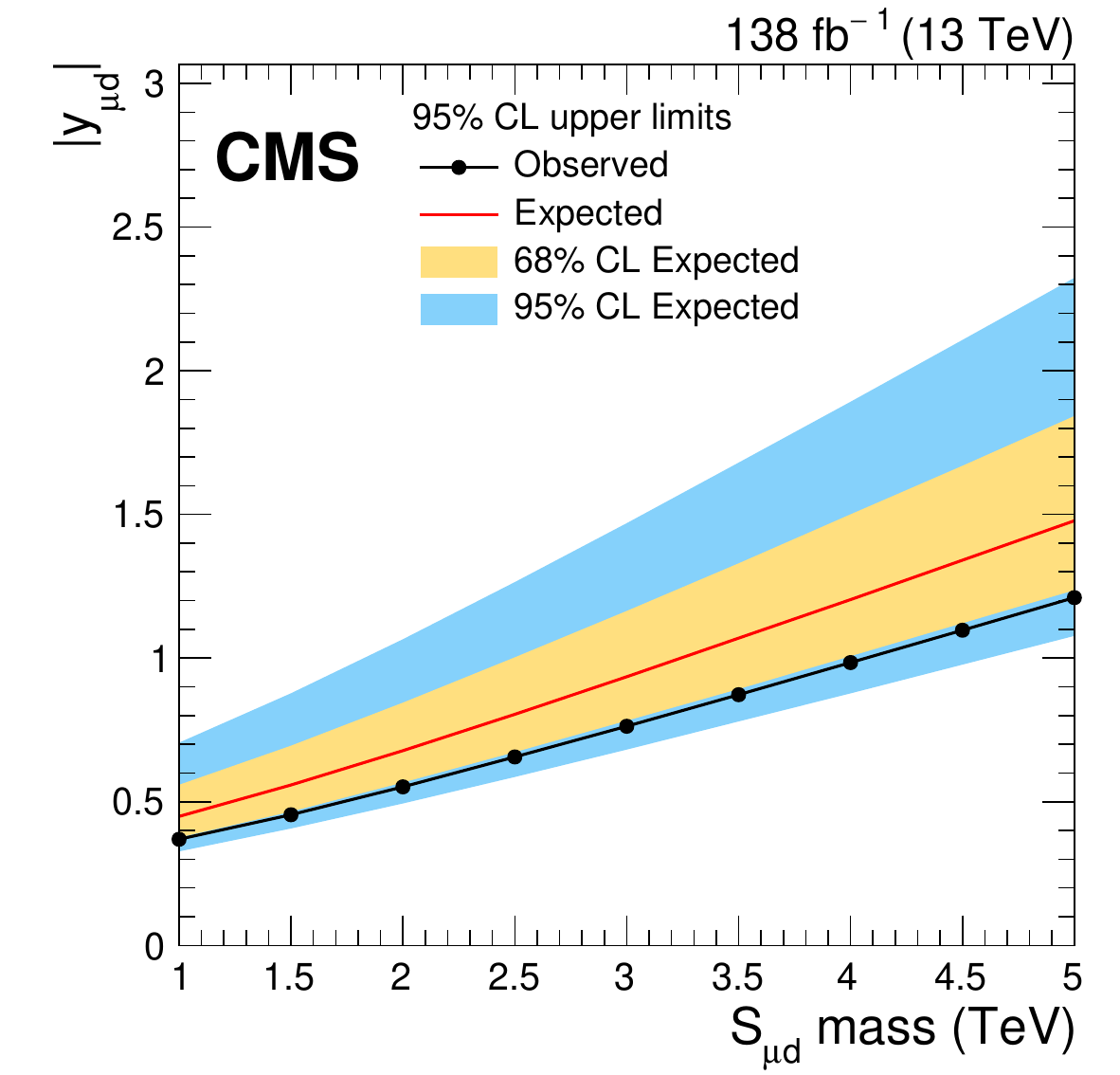}
  \caption{
    Upper limits at 95\%~\CL on the LQ-fermion couplings, $\abs{\ymu}$ (left) and $\abs{\ymd}$ (right), as a function of \mlq for scalar LQs coupled to muons. 
    The black points show the observed limits, the red line shows the expected limits, and the yellow and blue bands show the variations on the expected limit at 68\% and 95\% \CL, respectively. 
    }
  \label{fig:lims2}
\end{figure}

\begin{figure}[hbtp]
  \includegraphics[width = 0.49 \textwidth]{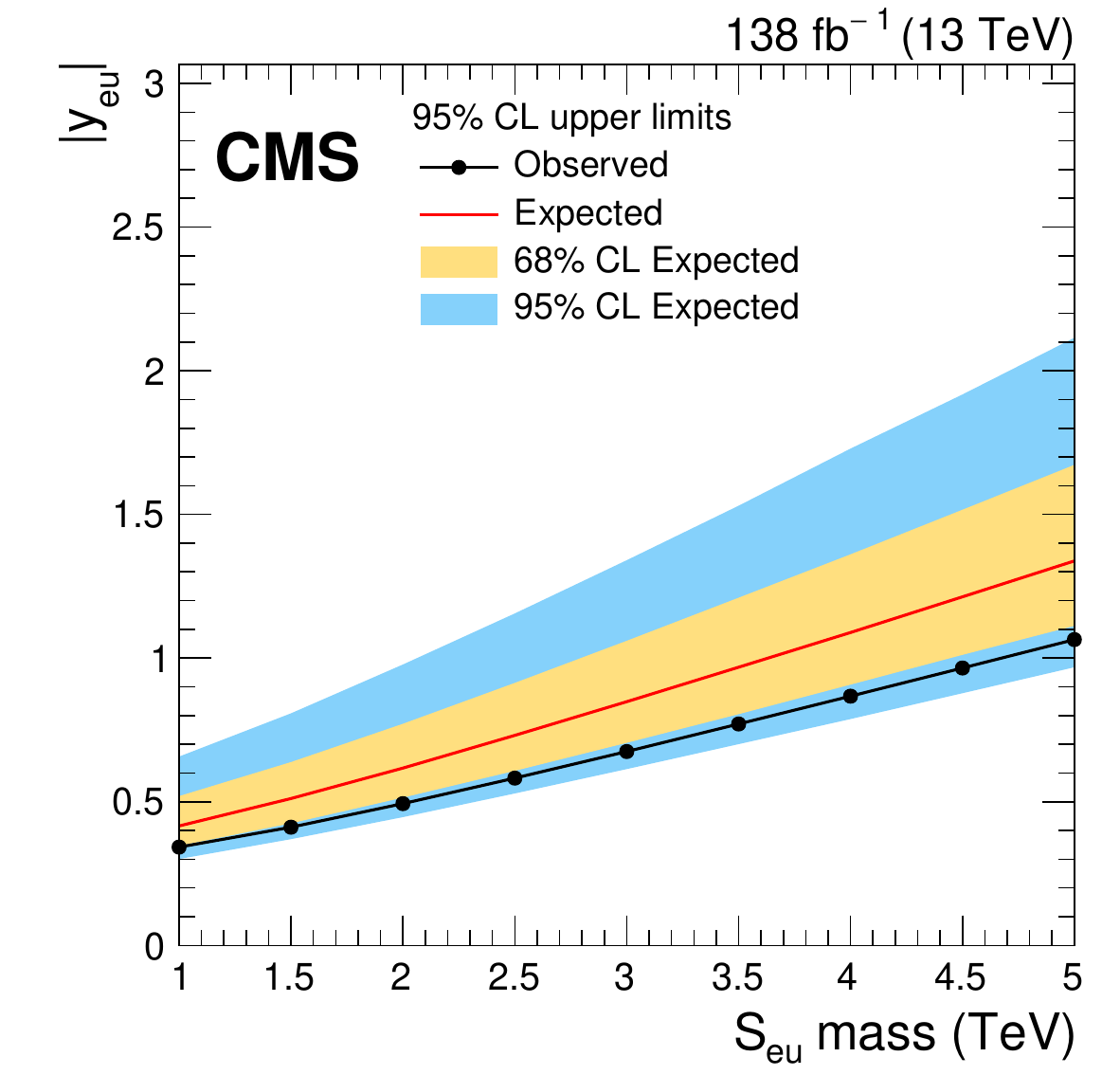}
  \includegraphics[width = 0.49 \textwidth]{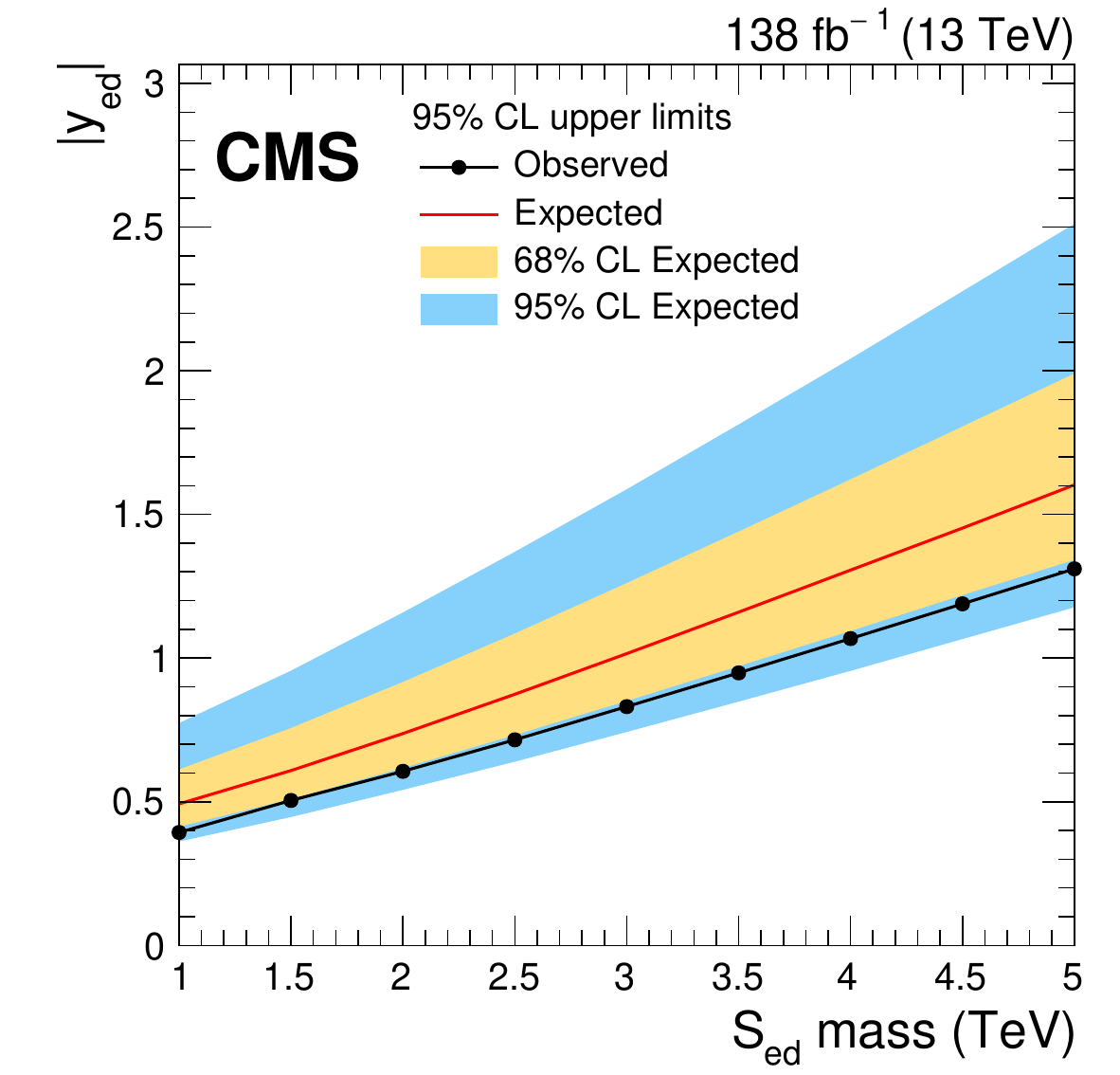}
  \caption{
    Upper limits at 95\%~\CL on the LQ-fermion couplings, $\abs{\yeu}$ (left) and $\abs{\yed}$ (right), as a function of \mlq for scalar LQs coupled to electrons.
    The black points show the observed limits, the red line shows the expected limits, and the yellow and blue bands show the variations on the expected limit at 68\% and 95\% \CL, respectively. 
    }
  \label{fig:lims1}
\end{figure}

\begin{figure}[hbtp]
  \includegraphics[width = 0.49 \textwidth]{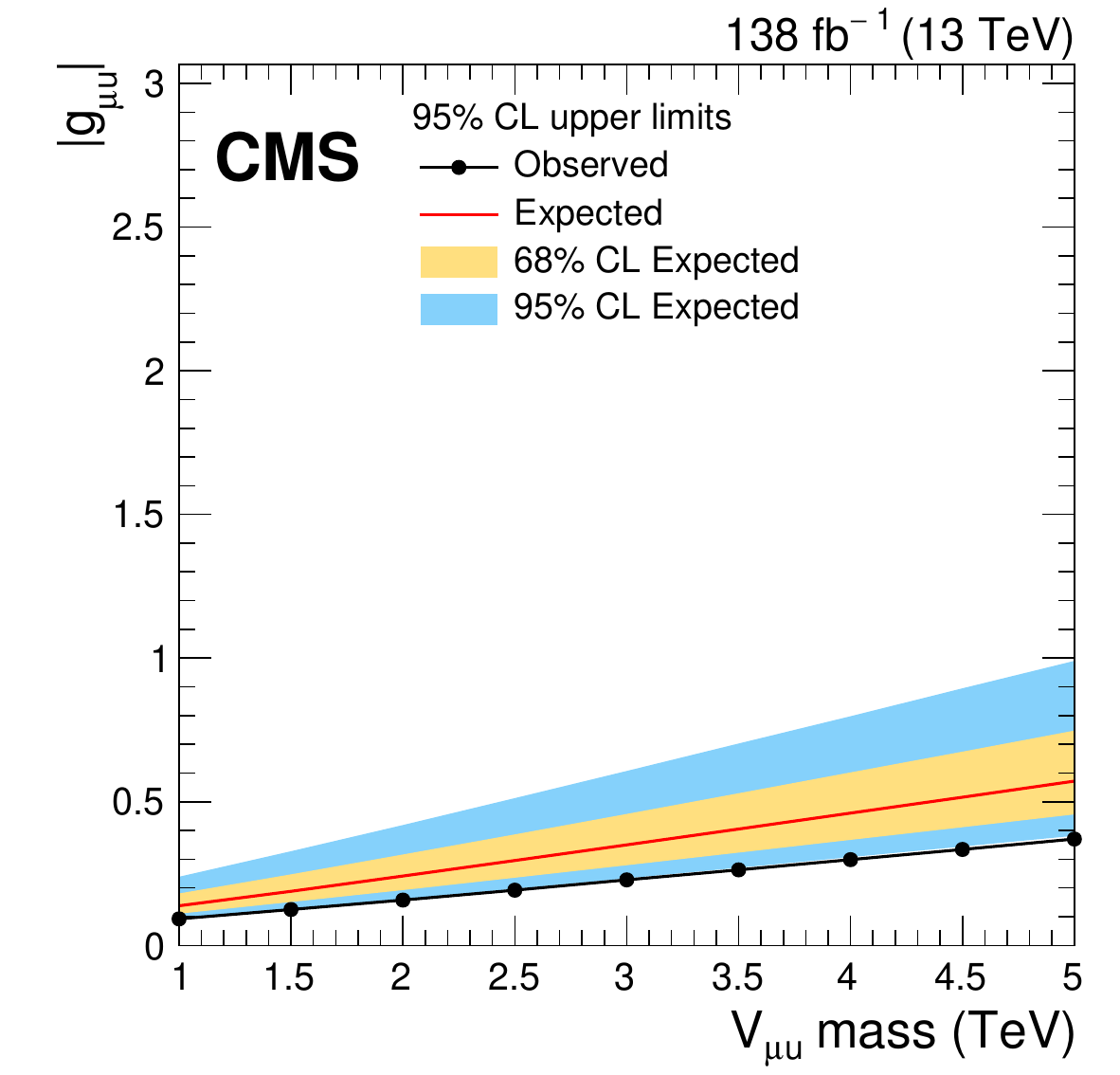}
  \includegraphics[width = 0.49 \textwidth]{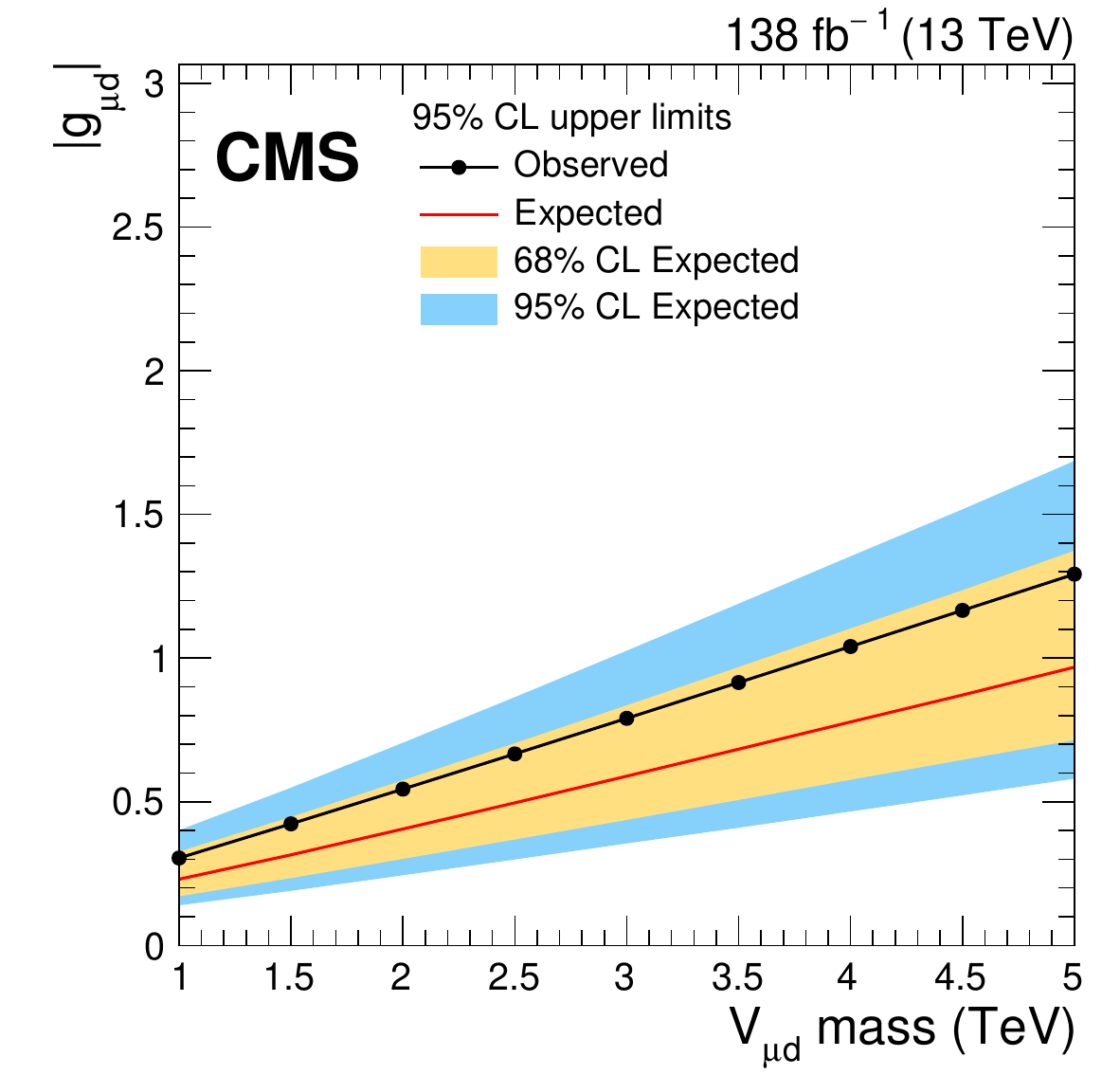}
  \caption{
    Upper limits at 95\%~\CL on the LQ-fermion couplings, $\abs{\gmu}$ (left) and $\abs{\gmd}$ (right), as a function of \mlq for vector LQs coupled to muons. 
    The black points show the observed limits, the red line shows the expected limits, and the yellow and blue bands show the variations on the expected limit at 68\% and 95\% \CL, respectively. 
    }
  \label{fig:lims4}
\end{figure}

\begin{figure}[hbtp]
  \includegraphics[width = 0.49 \textwidth]{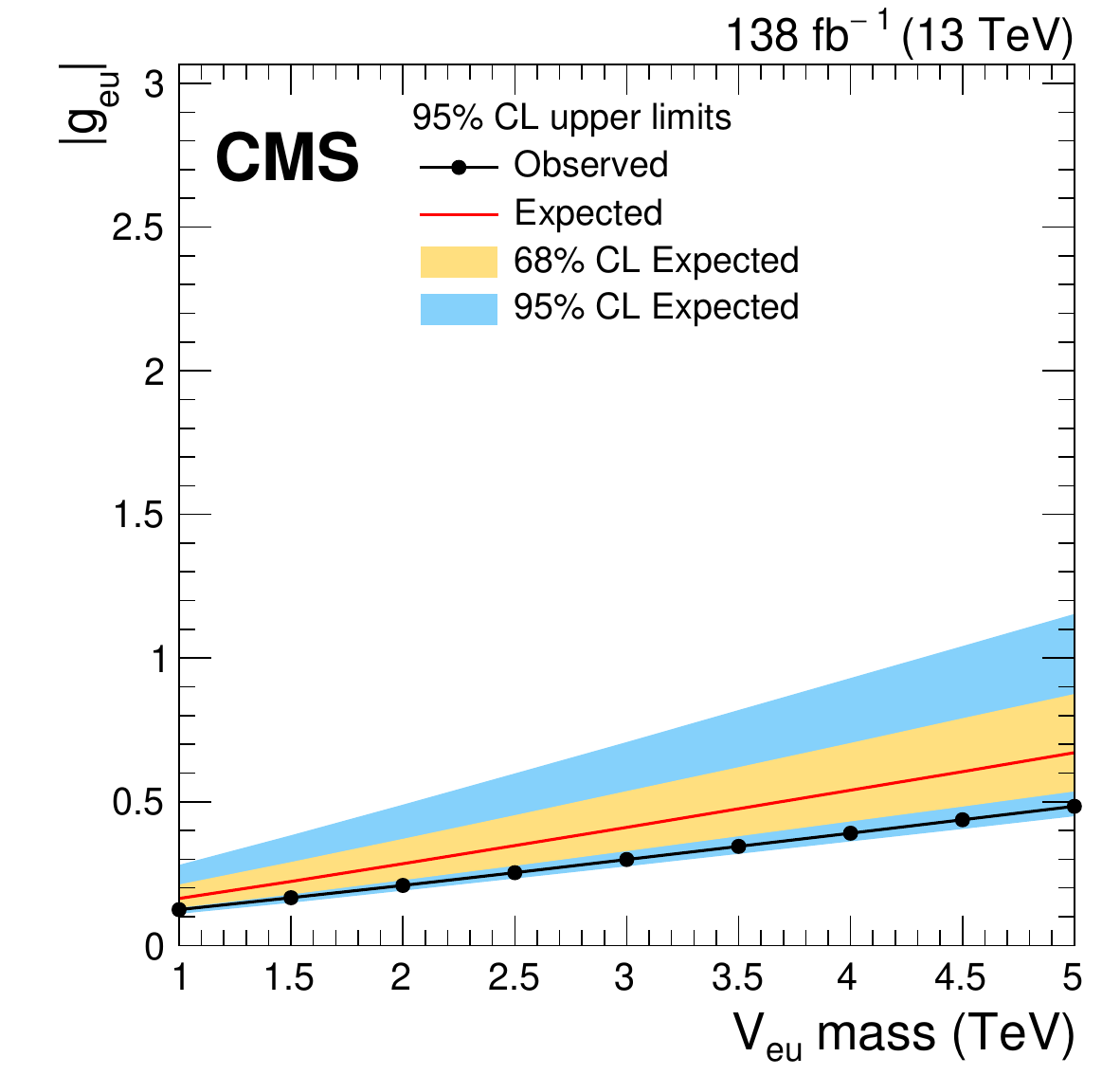}
  \includegraphics[width = 0.49 \textwidth]{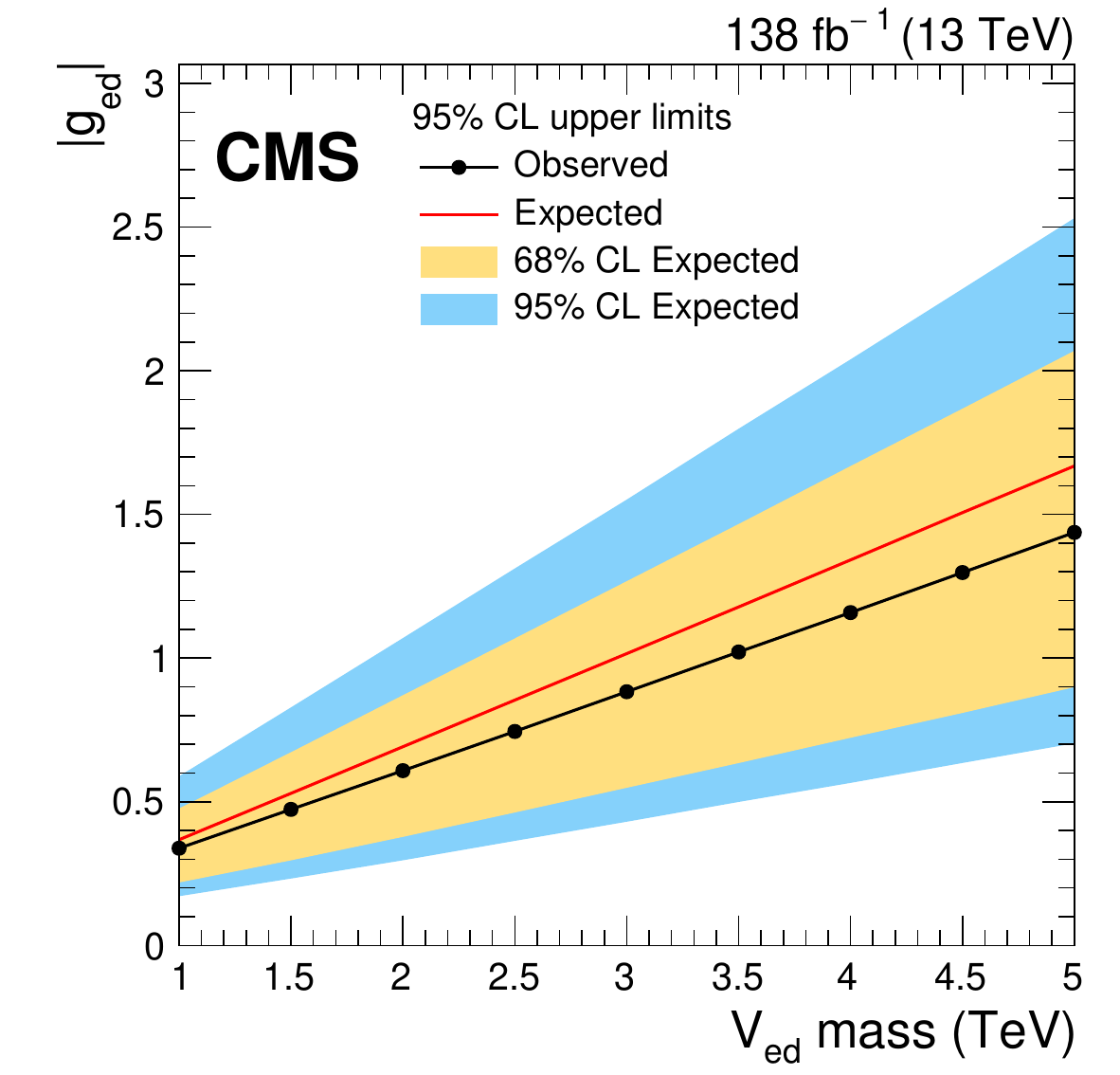}
  \caption{
    Upper limits at 95\%~\CL on the LQ-fermion couplings, $\abs{\geu}$ (left) and $\abs{\ged}$ (right), as a function of \mlq for vector LQs coupled to electrons. 
    The black points show the observed limits, the red line shows the expected limits, and the yellow and blue bands show the variations on the expected limit at 68\% and 95\% \CL, respectively. 
    }
  \label{fig:lims3}
\end{figure}

Limits at 95\% \CL are set on the absolute value of the LQ-fermion couplings as a function of the \mlq for each coupling and spin hypothesis. The limit-setting procedure employs a modified frequentist approach based on the \CLs criterion~\cite{CLS1,CLS2}, using an asymptotic approximation for the distribution of the test statistic~\cite{Cowan:2010js}. Limits extracted with this approach are nearly identical to those obtained from the Feldman--Cousins method.
The limits are shown in Figs.~\ref{fig:lims2} and \ref{fig:lims1} for the scalar case, and Figs.~\ref{fig:lims4} and \ref{fig:lims3} for the vector case, assuming branching fractions of 100\% for each channel. 

In general, for equal coupling values, the cross sections for vector LQ production are larger than for scalar LQ production, hence the limits on $\abs{\glq}$ are generally stronger than those on $\abs{\ylq}$. The limits on up-type couplings are stronger than the limits on down-type couplings because of the larger up quark content of the proton. In the $\PVed$ and $\PVmd$ cases, the interference process is negative, and the likelihood functions for these models are thus quite asymmetric and nonquadratic. This leads to smaller cross sections and weaker expected and observed limits than the other cases. The uncertainties in the expected limits are also accordingly larger than in the other cases. Further, the observed limit on $|\ged|$ is more stringent than the expected limit, despite the positive best fit value for the $\PVed$ mass of 2.5\TeV, as shown in Table~\ref{tab:results_vector}. This stems from the asymmetric nuisance impacts arising from the nonquadraticity in the likelihood, as discussed in Section~\ref{sec:sys_uncs}. 

At $\mlq \gg \mll$, the cross section scales as $\ylq^2/\mlq^2$ ($\glq^2/\mlq^2$), leading to the linearly increasing limits that are observed. Coupling values larger than $4\pi$ are nonperturbative.

For the scalar case, coupling values of 0.4--1.1 (0.3--1.0) are excluded for LQs coupled to electrons (muons) with masses between 1--5\TeV. 
These limits significantly extend the previous best limits from single production~\cite{CMS:2015xzc}, which excluded \PSeu and \PSed masses up to 1.7\TeV depending on the coupling, and pair production~\cite{CMS:2018ncu}, which excluded \PSeu and \PSed masses up to 1.4\TeV independent of the mass. 
Similarly, for the vector case, coupling values of 0.2--0.5 (0.1--0.4) are excluded for LQs coupled to electrons (muons). 
These are the first limits on vector LQs coupled to first- and second-generation fermions obtained via nonresonant $t$-channel exchange, extending the sensitivity of previous searches. 

\begin{table}[hbtp]
  \centering
  \topcaption{ Best fit values of \Azero, \Afour, and $\ylq^2$ for scalar LQ models. The Feldman--Cousins confidence interval for $\ylq^2$ is shown at 68\% \CL. Results are shown for a candidate \mlq of 2.5\TeV.}
  \renewcommand{\arraystretch}{1.4}
  \cmsTable{
  \begin{tabular}{l c c c c c}
      Model & \Azero & \Afour & $\ylq^2$ & Lower bound on $\ylq^2$ & Upper bound on $\ylq^2$ \\
      \hline
      \PSmu & $0.02 \pm 0.06$ &  $1.59 \pm 0.07$ & $-0.13_{-0.15}^{+0.14}$ (stat)$_{-0.11}^{+0.06}$ (syst) & 0 & 0.082 \\
      \PSmd & $0.02 \pm 0.06$ &  $1.60 \pm 0.07$ & $-0.11_{-0.20}^{+0.18}$ (stat)$_{-0.13}^{+0.09}$ (syst) & 0 & 0.119   \\
      \PSeu & $0.07 \pm 0.07$ & $1.61 \pm 0.08$ & $-0.10_{-0.17}^{+0.15}$ (stat)$_{-0.11}^{+0.07}$ (syst) & 0 & 0.093 \\
      \PSed & $0.07 \pm 0.07$ & $1.62 \pm 0.08$  & $ -0.09_{-0.23}^{+0.20}$ (stat)$_{-0.13}^{+0.11}$ (syst) & 0 & 0.138  \\
   \end{tabular}
 }
   \label{tab:results_scalar}
\end{table}

\begin{table}[hbtp]
  \centering
  \topcaption{ Best fit values of \Azero, \Afour, and $\glq^2$ for vector LQ models. The Feldman--Cousins confidence interval for $\glq^2$ is shown at 68\% \CL. Results are shown for a candidate \mlq of 2.5\TeV.}
  \renewcommand{\arraystretch}{1.4}
  \cmsTable{
      \begin{tabular}{l c c c c c}
        Model & \Azero & \Afour & $\glq^2$ & Lower bound on $\glq^2$ & Upper bound on $\glq^2$\\
        \hline
        \PVmu & 0.01 $\pm$ 0.05  & 1.63 $\pm$ 0.06 & $-0.10_{-0.02}^{+0.02}$ (stat)$_{-0.08}^{+0.04}$ (syst) & 0 & 0.029 \\
        \PVmd & 0.01 $\pm$ 0.05  & 1.61 $\pm$ 0.06 & $0.14_{-0.05}^{+0.05}$ (stat)$_{-0.07}^{+0.14}$ (syst) & 0.036 & 0.328 \\
        \PVeu & 0.05 $\pm$ 0.07  & 1.66 $\pm$ 0.08 & $-0.09_{-0.03}^{+0.03}$ (stat)$_{-0.08}^{+0.04}$ (syst) & 0 & 0.026 \\
        \PVed & 0.06 $\pm$ 0.07  & 1.64 $\pm$ 0.08 & $0.13_{-0.06}^{+0.06}$ (stat)$_{-0.09}^{+0.17}$ (syst) & 0.038 & 0.352 \\
         \end{tabular}
        \label{tab:results_vector}
      }
\end{table}

\section{Summary}\label{sec:summary}

A search for the $t$-channel exchange of leptoquarks (LQs) coupled to first- and second-gen\-eration fermions has been presented. 
The search uses proton-proton collision data at $\sqrt{s}=13\TeV$, corresponding to an integrated luminosity of 138\fbinv. 
Both scalar and vector LQs are considered, with masses between 1 and 5\TeV, for exclusive LQ couplings to an up or a down quark and a muon or an electron.
The $t$-channel exchange of an LQ modifies the dilepton angular distributions at masses well below the resonance mass of the LQ.
A template fit to the angular and invariant mass distributions of high-mass dilepton events is used to distinguish the signal process from the dominant Drell--Yan background, incorporating the interference effects.
No evidence for such LQs is observed. Limits are set at 95\% confidence level on the LQ-fermion couplings. 
For scalar LQs with masses in the 1--5\TeV range, coupling values greater than 0.4--1.3 (0.3--1.0) are excluded for LQs coupled to a muon or an electron, and an up (down) quark, and couplings greater than 0.3--1.4 (0.1--0.5) are excluded for the respective vector LQs.
This search is sensitive to LQs with significantly higher masses than prior single- and pair-production searches, establishing stringent limits on LQs with masses up to 5\TeV.

\begin{acknowledgments}
  We congratulate our colleagues in the CERN accelerator departments for the excellent performance of the LHC and thank the technical and administrative staffs at CERN and at other CMS institutes for their contributions to the success of the CMS effort. In addition, we gratefully acknowledge the computing centers and personnel of the Worldwide LHC Computing Grid and other centers for delivering so effectively the computing infrastructure essential to our analyses. Finally, we acknowledge the enduring support for the construction and operation of the LHC, the CMS detector, and the supporting computing infrastructure provided by the following funding agencies: SC (Armenia), BMBWF and FWF (Austria); FNRS and FWO (Belgium); CNPq, CAPES, FAPERJ, FAPERGS, and FAPESP (Brazil); MES and BNSF (Bulgaria); CERN; CAS, MoST, and NSFC (China); MINCIENCIAS (Colombia); MSES and CSF (Croatia); RIF (Cyprus); SENESCYT (Ecuador); ERC PRG, RVTT3 and MoER TK202 (Estonia); Academy of Finland, MEC, and HIP (Finland); CEA and CNRS/IN2P3 (France); SRNSF (Georgia); BMBF, DFG, and HGF (Germany); GSRI (Greece); NKFIH (Hungary); DAE and DST (India); IPM (Iran); SFI (Ireland); INFN (Italy); MSIP and NRF (Republic of Korea); MES (Latvia); LMTLT (Lithuania); MOE and UM (Malaysia); BUAP, CINVESTAV, CONACYT, LNS, SEP, and UASLP-FAI (Mexico); MOS (Montenegro); MBIE (New Zealand); PAEC (Pakistan); MES and NSC (Poland); FCT (Portugal); MESTD (Serbia); MICIU/AEI and PCTI (Spain); MOSTR (Sri Lanka); Swiss Funding Agencies (Switzerland); MST (Taipei); MHESI and NSTDA (Thailand); TUBITAK and TENMAK (Turkey); NASU (Ukraine); STFC (United Kingdom); DOE and NSF (USA).
  
  \hyphenation{Rachada-pisek} Individuals have received support from the Marie-Curie program and the European Research Council and Horizon 2020 Grant, contract Nos.\ 675440, 724704, 752730, 758316, 765710, 824093, 101115353, 101002207, and COST Action CA16108 (European Union); the Leventis Foundation; the Alfred P.\ Sloan Foundation; the Alexander von Humboldt Foundation; the Science Committee, project no. 22rl-037 (Armenia); the Fonds pour la Formation \`a la Recherche dans l'Industrie et dans l'Agriculture (FRIA-Belgium); the Beijing Municipal Science \& Technology Commission, No. Z191100007219010 and Fundamental Research Funds for the Central Universities (China); the Ministry of Education, Youth and Sports (MEYS) of the Czech Republic; the Shota Rustaveli National Science Foundation, grant FR-22-985 (Georgia); the Deutsche Forschungsgemeinschaft (DFG), among others, under Germany's Excellence Strategy -- EXC 2121 ``Quantum Universe" -- 390833306, and under project number 400140256 - GRK2497; the Hellenic Foundation for Research and Innovation (HFRI), Project Number 2288 (Greece); the Hungarian Academy of Sciences, the New National Excellence Program - \'UNKP, the NKFIH research grants K 131991, K 133046, K 138136, K 143460, K 143477, K 146913, K 146914, K 147048, 2020-2.2.1-ED-2021-00181, TKP2021-NKTA-64, and 2021-4.1.2-NEMZ\_KI-2024-00036 (Hungary); the Council of Science and Industrial Research, India; ICSC -- National Research Center for High Performance Computing, Big Data and Quantum Computing and FAIR -- Future Artificial Intelligence Research, funded by the NextGenerationEU program (Italy); the Latvian Council of Science; the Ministry of Education and Science, project no. 2022/WK/14, and the National Science Center, contracts Opus 2021/41/B/ST2/01369 and 2021/43/B/ST2/01552 (Poland); the Funda\c{c}\~ao para a Ci\^encia e a Tecnologia, grant CEECIND/01334/2018 (Portugal); the National Priorities Research Program by Qatar National Research Fund; MICIU/AEI/10.13039/501100011033, ERDF/EU, "European Union NextGenerationEU/PRTR", and Programa Severo Ochoa del Principado de Asturias (Spain); the Chulalongkorn Academic into Its 2nd Century Project Advancement Project, and the National Science, Research and Innovation Fund via the Program Management Unit for Human Resources \& Institutional Development, Research and Innovation, grant B39G670016 (Thailand); the Kavli Foundation; the Nvidia Corporation; the SuperMicro Corporation; the Welch Foundation, contract C-1845; and the Weston Havens Foundation (USA).  
\end{acknowledgments}

\bibliography{auto_generated}

\providecommand{\href}[2]{#2}\begingroup\raggedright\begin{thebibliography}{10}%
\makeatletter
\providecommand{\hrefCMSnoop }[0]{\@secondoftwo}%
\makeatother
\providecommand{\doi}{\texttt{doi:}\begingroup \urlstyle{tt}\Url}

\bibitem{Pati:1974yy}
\hrefCMSnoop {}{J.~C. Pati and A.~Salam, ``Lepton number as the fourth
  color'',} \textit{ Phys. Rev. D} \textbf{ 10} (1974) 275,
  \href{http://dx.doi.org/10.1103/PhysRevD.10.275}{\doi{10.1103/PhysRevD.10.275}}.
  [Erratum: \DOI{10.1103/PhysRevD.11.703.2}].

\bibitem{Georgi:1974sy}
\hrefCMSnoop {}{H.~Georgi and S.~L. Glashow, ``Unity of all elementary particle
  forces'',} \textit{ Phys. Rev. Lett.} \textbf{ 32} (1974) 438,
  \href{http://dx.doi.org/10.1103/PhysRevLett.32.438}{\doi{10.1103/PhysRevLett.32.438}}.

\bibitem{Senjanovic:1982ex}
\hrefCMSnoop {}{G.~Senjanovic and A.~Sokorac, ``{Light Leptoquarks in
  SO(10)}'',} \textit{ Z. Phys. C} \textbf{ 20} (1983) 255,
  \href{http://dx.doi.org/10.1007/BF01574858}{\doi{10.1007/BF01574858}}.

\bibitem{Davidson:2010uu}
\hrefCMSnoop {}{S.~Davidson and S.~Descotes-Genon, ``Minimal flavour violation
  for leptoquarks'',} \textit{ JHEP} \textbf{ 11} (2010) 073,
  \href{http://dx.doi.org/10.1007/JHEP11(2010)073}{\doi{10.1007/JHEP11(2010)073}},
  \href{http://www.arXiv.org/abs/1009.1998}{\texttt{arXiv:1009.1998}}.

\bibitem{Dorsner:2005fq}
\hrefCMSnoop {}{I.~Dorsner and P.~Fileviez~Perez, ``Unification without
  supersymmetry: Neutrino mass, proton decay and light leptoquarks'',} \textit{
  Nucl. Phys. B} \textbf{ 723} (2005) 53,
  \href{http://dx.doi.org/10.1016/j.nuclphysb.2005.06.016}{\doi{10.1016/j.nuclphysb.2005.06.016}},
  \href{http://www.arXiv.org/abs/hep-ph/0504276}{\texttt{arXiv:hep-ph/0504276}}.

\bibitem{Buchmuller:1986iq}
\hrefCMSnoop {}{W.~Buchmuller and D.~Wyler, ``Constraints on {SU(5)} type
  leptoquarks'',} \textit{ Phys. Lett. B} \textbf{ 177} (1986) 377,
  \href{http://dx.doi.org/10.1016/0370-2693(86)90771-9}{\doi{10.1016/0370-2693(86)90771-9}}.

\bibitem{Gripaios:2014tna}
\hrefCMSnoop {}{B.~Gripaios, M.~Nardecchia, and S.~A. Renner, ``Composite
  leptoquarks and anomalies in $b$-meson decays'',} \textit{ JHEP} \textbf{ 05}
  (2015) 006,
  \href{http://dx.doi.org/10.1007/JHEP05(2015)006}{\doi{10.1007/JHEP05(2015)006}},
  \href{http://www.arXiv.org/abs/1412.1791}{\texttt{arXiv:1412.1791}}.

\bibitem{Gripaios:2009dq}
\hrefCMSnoop {}{B.~Gripaios, ``Composite leptoquarks at the {LHC}'',} \textit{
  JHEP} \textbf{ 02} (2010) 045,
  \href{http://dx.doi.org/10.1007/JHEP02(2010)045}{\doi{10.1007/JHEP02(2010)045}},
  \href{http://www.arXiv.org/abs/0910.1789}{\texttt{arXiv:0910.1789}}.

\bibitem{Farhi:1980xs}
\hrefCMSnoop {}{E.~Farhi and L.~Susskind, ``Technicolor'',} \textit{ Phys.
  Rept.} \textbf{ 74} (1981) 277,
  \href{http://dx.doi.org/10.1016/0370-1573(81)90173-3}{\doi{10.1016/0370-1573(81)90173-3}}.

\bibitem{Dimopoulos:1979sp}
\hrefCMSnoop {}{S.~Dimopoulos, ``Technicolored signatures'',} \textit{ Nucl.
  Phys. B} \textbf{ 168} (1980) 69,
  \href{http://dx.doi.org/10.1016/0550-3213(80)90277-1}{\doi{10.1016/0550-3213(80)90277-1}}.

\bibitem{Schrempp:1984nj}
\hrefCMSnoop {}{B.~Schrempp and F.~Schrempp, ``Light leptoquarks'',} \textit{
  Phys. Lett. B} \textbf{ 153} (1985) 101,
  \href{http://dx.doi.org/10.1016/0370-2693(85)91450-9}{\doi{10.1016/0370-2693(85)91450-9}}.

\bibitem{Baker:2015qna}
\hrefCMSnoop {}{M.~J. Baker { et~al.}, ``The coannihilation codex'',} \textit{
  JHEP} \textbf{ 12} (2015) 120,
  \href{http://dx.doi.org/10.1007/JHEP12(2015)120}{\doi{10.1007/JHEP12(2015)120}},
  \href{http://www.arXiv.org/abs/1510.03434}{\texttt{arXiv:1510.03434}}.

\bibitem{Choi:2018stw}
\hrefCMSnoop {}{S.-M. Choi, Y.-J. Kang, H.~M. Lee, and T.-G. Ro,
  ``{L}epto-{Q}uark portal dark matter'',} \textit{ JHEP} \textbf{ 10} (2018)
  104,
  \href{http://dx.doi.org/10.1007/JHEP10(2018)104}{\doi{10.1007/JHEP10(2018)104}},
  \href{http://www.arXiv.org/abs/1807.06547}{\texttt{arXiv:1807.06547}}.

\bibitem{Barbier:2004ez}
\hrefCMSnoop {}{R.~Barbier { et~al.}, ``{R}-parity violating supersymmetry'',}
  \textit{ Phys. Rept.} \textbf{ 420} (2005) 1,
  \href{http://dx.doi.org/10.1016/j.physrep.2005.08.006}{\doi{10.1016/j.physrep.2005.08.006}},
  \href{http://www.arXiv.org/abs/hep-ph/0406039}{\texttt{arXiv:hep-ph/0406039}}.

\bibitem{Muong-2:2021ojo}
\hrefCMSnoop {}{{{Muon $g-2$}} Collaboration, ``Measurement of the positive
  muon anomalous magnetic moment to 0.46 ppm'',} \textit{ Phys. Rev. Lett.}
  \textbf{ 126} (2021) 141801,
  \href{http://dx.doi.org/10.1103/PhysRevLett.126.141801}{\doi{10.1103/PhysRevLett.126.141801}},
  \href{http://www.arXiv.org/abs/2104.03281}{\texttt{arXiv:2104.03281}}.

\bibitem{Borsanyi:2020mff}
\hrefCMSnoop {}{S.~Borsanyi { et~al.}, ``Leading hadronic contribution to the
  muon magnetic moment from lattice {QCD}'',} \textit{ Nature} \textbf{ 593}
  (2021) 51,
  \href{http://dx.doi.org/10.1038/s41586-021-03418-1}{\doi{10.1038/s41586-021-03418-1}},
  \href{http://www.arXiv.org/abs/2002.12347}{\texttt{arXiv:2002.12347}}.

\bibitem{Dorsner:2019itg}
\hrefCMSnoop {}{I.~Dor{\v{s}}ner, S.~Fajfer, and O.~Sumensari, ``{Muon $g-2$
  and scalar leptoquark mixing}'',} \textit{ JHEP} \textbf{ 06} (2020) 089,
  \href{http://dx.doi.org/10.1007/JHEP06(2020)089}{\doi{10.1007/JHEP06(2020)089}},
  \href{http://www.arXiv.org/abs/1910.03877}{\texttt{arXiv:1910.03877}}.

\bibitem{Muong-2:2025xyk}
\hrefCMSnoop {}{{{Muon $g-2$}} Collaboration, ``Measurement of the positive
  muon anomalous magnetic moment to 127 ppb'',} 2025.
  \href{http://www.arXiv.org/abs/2506.03069}{\texttt{arXiv:2506.03069}}.

\bibitem{Dorsner:2016wpm}
I.~Dor{\v{s}}ner\hrefCMSnoop {}{ { et~al.}, ``Physics of leptoquarks in
  precision experiments and at particle colliders'',} \textit{ Phys. Rept.}
  \textbf{ 641} (2016) 1,
  \href{http://dx.doi.org/10.1016/j.physrep.2016.06.001}{\doi{10.1016/j.physrep.2016.06.001}},
  \href{http://www.arXiv.org/abs/1603.04993}{\texttt{arXiv:1603.04993}}.

\bibitem{CMS:2023qdw}
\hrefCMSnoop {}{{CMS Collaboration}, ``{Search for a third-generation
  leptoquark coupled to a \ensuremath{\tau} lepton and a b quark through
  single, pair, and nonresonant production in proton-proton collisions at $
  \sqrt{s} $ = 13 TeV}'',} \textit{ JHEP} \textbf{ 05} (2024) 311,
  \href{http://dx.doi.org/10.1007/JHEP05(2024)311}{\doi{10.1007/JHEP05(2024)311}},
  \href{http://www.arXiv.org/abs/2308.07826}{\texttt{arXiv:2308.07826}}.

\bibitem{ATLAS:2023uox}
\hrefCMSnoop {}{{ATLAS Collaboration}, ``Search for pair production of
  third-generation leptoquarks decaying into a bottom quark and a $\tau
  $-lepton with the {ATLAS} detector'',} \textit{ Eur. Phys. J. C} \textbf{ 83}
  (2023) 1075,
  \href{http://dx.doi.org/10.1140/epjc/s10052-023-12104-7}{\doi{10.1140/epjc/s10052-023-12104-7}},
  \href{http://www.arXiv.org/abs/2303.01294}{\texttt{arXiv:2303.01294}}.

\bibitem{CMS:2023arc}
\hrefCMSnoop {}{{CMS Collaboration}, ``{Searches for additional Higgs bosons
  and for vector leptoquarks in $\tau\tau$ final states in proton-proton
  collisions at $\sqrt{s}$ = 13 TeV}'',} \textit{ JHEP} \textbf{ 07} (2023)
  073,
  \href{http://dx.doi.org/10.1007/JHEP07(2023)073}{\doi{10.1007/JHEP07(2023)073}},
  \href{http://www.arXiv.org/abs/2208.02717}{\texttt{arXiv:2208.02717}}.

\bibitem{ATLAS:2020dsk}
\hrefCMSnoop {}{{ATLAS Collaboration}, ``{Search for pairs of scalar
  leptoquarks decaying into quarks and electrons or muons in $ \sqrt{s} $ = 13
  TeV $pp$ collisions with the ATLAS detector}'',} \textit{ JHEP} \textbf{ 10}
  (2020) 112,
  \href{http://dx.doi.org/10.1007/JHEP10(2020)112}{\doi{10.1007/JHEP10(2020)112}},
  \href{http://www.arXiv.org/abs/2006.05872}{\texttt{arXiv:2006.05872}}.

\bibitem{CMS:2015xzc}
\hrefCMSnoop {}{{CMS Collaboration}, ``Search for single production of scalar
  leptoquarks in proton-proton collisions at $\sqrt{s} = 8$ {TeV}'',} \textit{
  Phys. Rev. D} \textbf{ 93} (2016) 032005,
  \href{http://dx.doi.org/10.1103/PhysRevD.93.032005}{\doi{10.1103/PhysRevD.93.032005}},
  \href{http://www.arXiv.org/abs/1509.03750}{\texttt{arXiv:1509.03750}}.
  [Erratum: \DOI{10.1103/PhysRevD.95.039906}].

\bibitem{CMS:2018ncu}
\hrefCMSnoop {}{{CMS Collaboration}, ``Search for pair production of
  first-generation scalar leptoquarks at $\sqrt{s} =$ 13 {TeV}'',} \textit{
  Phys. Rev. D} \textbf{ 99} (2019) 052002,
  \href{http://dx.doi.org/10.1103/PhysRevD.99.052002}{\doi{10.1103/PhysRevD.99.052002}},
  \href{http://www.arXiv.org/abs/1811.01197}{\texttt{arXiv:1811.01197}}.

\bibitem{CMS:2018lab}
\hrefCMSnoop {}{{CMS Collaboration}, ``Search for pair production of
  second-generation leptoquarks at $\sqrt{s}=$ 13 {TeV}'',} \textit{ Phys. Rev.
  D} \textbf{ 99} (2019) 032014,
  \href{http://dx.doi.org/10.1103/PhysRevD.99.032014}{\doi{10.1103/PhysRevD.99.032014}},
  \href{http://www.arXiv.org/abs/1808.05082}{\texttt{arXiv:1808.05082}}.

\bibitem{CMS:2015gua}
\hrefCMSnoop {}{{CMS Collaboration}, ``{Search for pair production of first and
  second generation leptoquarks in proton-proton collisions at $\sqrt{s}$ = 8
  TeV}'',} \textit{ Phys. Rev. D} \textbf{ 93} (2016) 032004,
  \href{http://dx.doi.org/10.1103/PhysRevD.93.032004}{\doi{10.1103/PhysRevD.93.032004}},
  \href{http://www.arXiv.org/abs/1509.03744}{\texttt{arXiv:1509.03744}}.

\bibitem{Raj:2016aky}
\hrefCMSnoop {}{N.~Raj, ``Anticipating nonresonant new physics in dilepton
  angular spectra at the {LHC}'',} \textit{ Phys. Rev. D} \textbf{ 95} (2017)
  015011,
  \href{http://dx.doi.org/10.1103/PhysRevD.95.015011}{\doi{10.1103/PhysRevD.95.015011}},
  \href{http://www.arXiv.org/abs/1610.03795}{\texttt{arXiv:1610.03795}}.

\bibitem{Crivellin:2021egp}
\hrefCMSnoop {}{A.~Crivellin, D.~M{\"u}ller, and L.~Schnell, ``Combined
  constraints on first generation leptoquarks'',} \textit{ Phys. Rev. D}
  \textbf{ 103} (2021) 115023,
  \href{http://dx.doi.org/10.1103/PhysRevD.103.115023}{\doi{10.1103/PhysRevD.103.115023}},
  \href{http://www.arXiv.org/abs/2104.06417}{\texttt{arXiv:2104.06417}}.
  [Addendum: \DOI{10.1103/PhysRevD.104.055020}].

\bibitem{Bigaran:2020jil_r2}
\hrefCMSnoop {}{I.~Bigaran and R.~R. Volkas, ``Getting chirality right: Single
  scalar leptoquark solutions to the $(g-2)_{e,\mu}$ puzzle'',} \textit{ Phys.
  Rev. D} \textbf{ 102} (2020) 075037,
  \href{http://dx.doi.org/10.1103/PhysRevD.102.075037}{\doi{10.1103/PhysRevD.102.075037}},
  \href{http://www.arXiv.org/abs/2002.12544}{\texttt{arXiv:2002.12544}}.

\bibitem{Dorsner:2023ykm}
\hrefCMSnoop {}{I.~Dorsner, ``{Novel leptoquark pair production @LHC}'',}
  \textit{ PoS} \textbf{ CORFU2022} (2023) 111,
  \href{http://dx.doi.org/10.22323/1.436.0111}{\doi{10.22323/1.436.0111}}.

\bibitem{Robinson:2018gza}
\hrefCMSnoop {}{D.~J. Robinson, B.~Shakya, and J.~Zupan, ``{Right-handed
  neutrinos and R(D$^{*}$)}'',} \textit{ JHEP} \textbf{ 02} (2019) 119,
  \href{http://dx.doi.org/10.1007/JHEP02(2019)119}{\doi{10.1007/JHEP02(2019)119}},
  \href{http://www.arXiv.org/abs/1807.04753}{\texttt{arXiv:1807.04753}}.

\bibitem{Becirevic:2016oho_u3}
\hrefCMSnoop {}{D.~Be{\v{c}}irevi{\'c}, N.~Ko{\v{s}}nik, O.~Sumensari, and
  R.~Zukanovich~Funchal, ``Palatable leptoquark scenarios for lepton flavor
  violation in exclusive $b\to s\ell_1\ell_2$ modes'',} \textit{ JHEP} \textbf{
  11} (2016) 035,
  \href{http://dx.doi.org/10.1007/JHEP11(2016)035}{\doi{10.1007/JHEP11(2016)035}},
  \href{http://www.arXiv.org/abs/1608.07583}{\texttt{arXiv:1608.07583}}.

\bibitem{HEPData}
\hrefCMSnoop {}{}{HEPD}ata record for this analysis, 2024.
\newblock
  \href{http://dx.doi.org/10.17182/hepdata.156848}{\doi{10.17182/hepdata.156848}}.

\bibitem{Collins:1977iv}
\hrefCMSnoop {}{J.~C. Collins and D.~E. Soper, ``{Angular distribution of
  dileptons in high-energy hadron collisions}'',} \textit{ Phys. Rev. D}
  \textbf{ 16} (1977) 2219,
  \href{http://dx.doi.org/10.1103/PhysRevD.16.2219}{\doi{10.1103/PhysRevD.16.2219}}.

\bibitem{CMS:2022uul}
\hrefCMSnoop {}{{CMS Collaboration}, ``Measurement of the {Drell-Yan}
  forward-backward asymmetry at high dilepton masses in proton-proton
  collisions at $\sqrt{s}$ = 13 {TeV}'',} \textit{ JHEP} \textbf{ 08} (2022)
  063,
  \href{http://dx.doi.org/10.1007/JHEP08(2022)063}{\doi{10.1007/JHEP08(2022)063}},
  \href{http://www.arXiv.org/abs/2202.12327}{\texttt{arXiv:2202.12327}}.

\bibitem{Mirkes:1992hu}
\hrefCMSnoop {}{E.~Mirkes, ``Angular decay distribution of leptons from {W}
  bosons at {NLO} in hadronic collisions'',} \textit{ Nucl. Phys. B} \textbf{
  387} (1992) 3,
  \href{http://dx.doi.org/10.1016/0550-3213(92)90046-E}{\doi{10.1016/0550-3213(92)90046-E}}.

\bibitem{Mirkes:1994dp}
\hrefCMSnoop {}{E.~Mirkes and J.~Ohnemus, ``Angular distributions of
  {Drell-Yan} lepton pairs at the {Tevatron: Order $\alpha-s^{2}$} corrections
  and {Monte Carlo} studies'',} \textit{ Phys. Rev. D} \textbf{ 51} (1995)
  4891,
  \href{http://dx.doi.org/10.1103/PhysRevD.51.4891}{\doi{10.1103/PhysRevD.51.4891}},
  \href{http://www.arXiv.org/abs/hep-ph/9412289}{\texttt{arXiv:hep-ph/9412289}}.

\bibitem{CMS:2008xjf}
\hrefCMSnoop {}{{CMS Collaboration}, ``The {CMS} experiment at the {CERN
  LHC}'',} \textit{ JINST} \textbf{ 3} (2008) S08004,
  \href{http://dx.doi.org/10.1088/1748-0221/3/08/S08004}{\doi{10.1088/1748-0221/3/08/S08004}}.

\bibitem{CMS:2023gfb}
\hrefCMSnoop {}{{CMS Collaboration}, ``Development of the {CMS} detector for
  the {CERN LHC Run 3}'',} \textit{ JINST} \textbf{ 19} (2024) P05064,
  \href{http://dx.doi.org/10.1088/1748-0221/19/05/P05064}{\doi{10.1088/1748-0221/19/05/P05064}},
  \href{http://www.arXiv.org/abs/2309.05466}{\texttt{arXiv:2309.05466}}.

\bibitem{Sirunyan:2020zal}
\hrefCMSnoop {}{{CMS Collaboration}, ``Performance of the {CMS Level-1} trigger
  in proton-proton collisions at {$\sqrt{s} = 13$\,TeV}'',} \textit{ JINST}
  \textbf{ 15} (2020) P10017,
  \href{http://dx.doi.org/10.1088/1748-0221/15/10/P10017}{\doi{10.1088/1748-0221/15/10/P10017}},
  \href{http://www.arXiv.org/abs/2006.10165}{\texttt{arXiv:2006.10165}}.

\bibitem{Khachatryan:2016bia}
\hrefCMSnoop {}{{CMS Collaboration}, ``The {CMS} trigger system'',} \textit{
  JINST} \textbf{ 12} (2017) P01020,
  \href{http://dx.doi.org/10.1088/1748-0221/12/01/P01020}{\doi{10.1088/1748-0221/12/01/P01020}},
\href{http://www.arXiv.org/abs/1609.02366}{\texttt{arXiv:1609.02366}}.

\bibitem{Sirunyan:2017ulk}
\hrefCMSnoop {}{{CMS Collaboration}, ``Particle-flow reconstruction and global
  event description with the {CMS} detector'',} \textit{ JINST} \textbf{ 12}
  (2017) P10003,
  \href{http://dx.doi.org/10.1088/1748-0221/12/10/P10003}{\doi{10.1088/1748-0221/12/10/P10003}},
\href{http://www.arXiv.org/abs/1706.04965}{\texttt{arXiv:1706.04965}}.

\bibitem{CMS-TDR-15-02}
\href {http://cds.cern.ch/record/2020886}{{CMS Collaboration}, ``Technical
  proposal for the {Phase-II} upgrade of the {Compact Muon Solenoid}'',} CMS
  Technical Proposal CERN-LHCC-2015-010, CMS-TDR-15-02, 2015.

\bibitem{CMS:2018rym}
\hrefCMSnoop {}{{CMS Collaboration}, ``Performance of the {CMS} muon detector
  and muon reconstruction with proton-proton collisions at $\sqrt{s}=$ 13
  {TeV}'',} \textit{ JINST} \textbf{ 13} (2018) P06015,
  \href{http://dx.doi.org/10.1088/1748-0221/13/06/P06015}{\doi{10.1088/1748-0221/13/06/P06015}},
  \href{http://www.arXiv.org/abs/1804.04528}{\texttt{arXiv:1804.04528}}.

\bibitem{CMS:2020uim}
\hrefCMSnoop {}{{CMS Collaboration}, ``Electron and photon reconstruction and
  identification with the {CMS} experiment at the {CERN LHC}'',} \textit{
  JINST} \textbf{ 16} (2021) P05014,
  \href{http://dx.doi.org/10.1088/1748-0221/16/05/P05014}{\doi{10.1088/1748-0221/16/05/P05014}},
  \href{http://www.arXiv.org/abs/2012.06888}{\texttt{arXiv:2012.06888}}.

\bibitem{Sirunyan_2020pileup}
\hrefCMSnoop {}{{CMS Collaboration}, ``Pileup mitigation at {CMS} in 13 {TeV}
  data'',} \textit{ JINST} \textbf{ 15} (2020) P09018,
  \href{http://dx.doi.org/10.1088/1748-0221/15/09/p09018}{\doi{10.1088/1748-0221/15/09/p09018}},
  \href{http://www.arXiv.org/abs/2003.00503}{\texttt{arXiv:2003.00503}}.

\bibitem{Alwall:2014hca}
J.~Alwall\hrefCMSnoop {}{ { et~al.}, ``The automated computation of tree-level
  and next-to-leading order differential cross sections, and their matching to
  parton shower simulations'',} \textit{ JHEP} \textbf{ 07} (2014) 079,
  \href{http://dx.doi.org/10.1007/JHEP07(2014)079}{\doi{10.1007/JHEP07(2014)079}},
\href{http://www.arXiv.org/abs/1405.0301}{\texttt{arXiv:1405.0301}}.

\bibitem{Sjostrand:2014zea}
T.~Sj{\"o}strand\hrefCMSnoop {}{ { et~al.}, ``An introduction to {PYTHIA
  8.2}'',} \textit{ Comput. Phys. Commun.} \textbf{ 191} (2015) 159,
  \href{http://dx.doi.org/10.1016/j.cpc.2015.01.024}{\doi{10.1016/j.cpc.2015.01.024}},
\href{http://www.arXiv.org/abs/1410.3012}{\texttt{arXiv:1410.3012}}.

\bibitem{CMS:2019csb}
\hrefCMSnoop {}{{CMS Collaboration}, ``Extraction and validation of a new set
  of {CMS PYTHIA8} tunes from underlying-event measurements'',} \textit{ Eur.
  Phys. J. C} \textbf{ 80} (2020) 4,
  \href{http://dx.doi.org/10.1140/epjc/s10052-019-7499-4}{\doi{10.1140/epjc/s10052-019-7499-4}},
  \href{http://www.arXiv.org/abs/1903.12179}{\texttt{arXiv:1903.12179}}.

\bibitem{CMS:2015wcf}
\hrefCMSnoop {}{{CMS Collaboration}, ``{Event generator tunes obtained from
  underlying event and multiparton scattering measurements}'',} \textit{ Eur.
  Phys. J. C} \textbf{ 76} (2016) 155,
  \href{http://dx.doi.org/10.1140/epjc/s10052-016-3988-x}{\doi{10.1140/epjc/s10052-016-3988-x}},
  \href{http://www.arXiv.org/abs/1512.00815}{\texttt{arXiv:1512.00815}}.

\bibitem{Ball:2010de}
R.~D. Ball\hrefCMSnoop {}{ { et~al.}, ``A first unbiased global {NLO}
  determination of parton distributions and their uncertainties'',} \textit{
  Nucl. Phys. B} \textbf{ 838} (2010) 136,
  \href{http://dx.doi.org/10.1016/j.nuclphysb.2010.05.008}{\doi{10.1016/j.nuclphysb.2010.05.008}},
\href{http://www.arXiv.org/abs/1002.4407}{\texttt{arXiv:1002.4407}}.

\bibitem{Ball:2014uwa}
\hrefCMSnoop {}{{NNPDF} Collaboration, ``Parton distributions for the {LHC Run
  II}'',} \textit{ JHEP} \textbf{ 04} (2015) 040,
  \href{http://dx.doi.org/10.1007/JHEP04(2015)040}{\doi{10.1007/JHEP04(2015)040}},
\href{http://www.arXiv.org/abs/1410.8849}{\texttt{arXiv:1410.8849}}.

\bibitem{Ball:2018iqk}
\hrefCMSnoop {}{{NNPDF} Collaboration, ``{Precision determination of the strong
  coupling constant within a global PDF analysis}'',} \textit{ Eur. Phys. J. C}
  \textbf{ 78} (2018) 408,
  \href{http://dx.doi.org/10.1140/epjc/s10052-018-5897-7}{\doi{10.1140/epjc/s10052-018-5897-7}},
  \href{http://www.arXiv.org/abs/1802.03398}{\texttt{arXiv:1802.03398}}.

\bibitem{Nason:2004rx}
\hrefCMSnoop {}{P.~Nason, ``A new method for combining {NLO QCD} with shower
  {M}onte {C}arlo algorithms'',} \textit{ JHEP} \textbf{ 11} (2004) 040,
  \href{http://dx.doi.org/10.1088/1126-6708/2004/11/040}{\doi{10.1088/1126-6708/2004/11/040}},
\href{http://www.arXiv.org/abs/hep-ph/0409146}{\texttt{arXiv:hep-ph/0409146}}.

\bibitem{Frixione:2007vw}
\hrefCMSnoop {}{S.~Frixione, P.~Nason, and C.~Oleari, ``Matching {NLO QCD}
  computations with parton shower simulations: the {POWHEG} method'',} \textit{
  JHEP} \textbf{ 11} (2007) 070,
  \href{http://dx.doi.org/10.1088/1126-6708/2007/11/070}{\doi{10.1088/1126-6708/2007/11/070}},
\href{http://www.arXiv.org/abs/0709.2092}{\texttt{arXiv:0709.2092}}.

\bibitem{Alioli:2010xd}
\hrefCMSnoop {}{S.~Alioli, P.~Nason, C.~Oleari, and E.~Re, ``A general
  framework for implementing {NLO} calculations in shower {Monte Carlo}
  programs: the {POWHEG BOX}'',} \textit{ JHEP} \textbf{ 06} (2010) 043,
  \href{http://dx.doi.org/10.1007/JHEP06(2010)043}{\doi{10.1007/JHEP06(2010)043}},
\href{http://www.arXiv.org/abs/1002.2581}{\texttt{arXiv:1002.2581}}.

\bibitem{Frixione:2007nw}
\hrefCMSnoop {}{S.~Frixione, P.~Nason, and G.~Ridolfi, ``A positive-weight
  next-to-leading-order {Monte Carlo} for heavy flavour hadroproduction'',}
  \textit{ JHEP} \textbf{ 09} (2007) 126,
  \href{http://dx.doi.org/10.1088/1126-6708/2007/09/126}{\doi{10.1088/1126-6708/2007/09/126}},
  \href{http://www.arXiv.org/abs/0707.3088}{\texttt{arXiv:0707.3088}}.

\bibitem{CMS-PAS-TOP-16-021}
\href {https://cds.cern.ch/record/2235192}{{CMS Collaboration},
  ``Investigations of the impact of the parton shower tuning in {Pythia8} in
  the modelling of $\mathrm{t\overline{t}}$ at $\sqrt{s}=8$ and 13 {TeV}'',}
  CMS Physics Analysis Summary CMS-PAS-TOP-16-021, 2016.

\bibitem{Melia:2011tj}
\hrefCMSnoop {}{T.~Melia, P.~Nason, R.~Rontsch, and G.~Zanderighi, ``{$W^+ W^-$
  , $W Z$ and $Z Z$ production in the POWHEG BOX}'',} \textit{ JHEP} \textbf{
  11} (2011) 078,
  \href{http://dx.doi.org/10.1007/JHEP11(2011)078}{\doi{10.1007/JHEP11(2011)078}},
  \href{http://www.arXiv.org/abs/1107.5051}{\texttt{arXiv:1107.5051}}.

\bibitem{Nason:2013ydw}
\hrefCMSnoop {}{P.~Nason and G.~Zanderighi, ``{$W^+ W^-$ , $W Z$ and $Z Z$
  production in the POWHEG-BOX-V2}'',} \textit{ Eur. Phys. J. C} \textbf{ 74}
  (2014) 2702,
  \href{http://dx.doi.org/10.1140/epjc/s10052-013-2702-5}{\doi{10.1140/epjc/s10052-013-2702-5}},
  \href{http://www.arXiv.org/abs/1311.1365}{\texttt{arXiv:1311.1365}}.

\bibitem{Forthomme:2018ecc}
\hrefCMSnoop {}{L.~Forthomme, ``{CepGen \textendash{} A generic central
  exclusive processes event generator for hadron-hadron collisions}'',}
  \textit{ Comput. Phys. Commun.} \textbf{ 271} (2022) 108225,
  \href{http://dx.doi.org/10.1016/j.cpc.2021.108225}{\doi{10.1016/j.cpc.2021.108225}},
  \href{http://www.arXiv.org/abs/1808.06059}{\texttt{arXiv:1808.06059}}.

\bibitem{Vermaseren:1982cz}
\hrefCMSnoop {}{J.~A.~M. Vermaseren, ``Two photon processes at very
  high-energies'',} \textit{ Nucl. Phys. B} \textbf{ 229} (1983) 347,
  \href{http://dx.doi.org/10.1016/0550-3213(83)90336-X}{\doi{10.1016/0550-3213(83)90336-X}}.

\bibitem{Baranov:1991yq}
\hrefCMSnoop {}{S.~Baranov, O.~D{\"u}enger, H.~Shooshtari, and J.~Vermaseren,
  ``{LPAIR: A generator for lepton pair production}'',} in \textit{ {Workshop
  on Physics at HERA}}, p.~1478.
\newblock 1991.

\bibitem{Sjostrand:2006za}
\hrefCMSnoop {}{T.~Sj{\"o}strand, S.~Mrenna, and P.~Z. Skands, ``{PYTHIA 6.4}
  physics and manual'',} \textit{ JHEP} \textbf{ 05} (2006) 026,
  \href{http://dx.doi.org/10.1088/1126-6708/2006/05/026}{\doi{10.1088/1126-6708/2006/05/026}},
  \href{http://www.arXiv.org/abs/hep-ph/0603175}{\texttt{arXiv:hep-ph/0603175}}.

\bibitem{Suri:1971yx}
\hrefCMSnoop {}{A.~Suri and D.~R. Yennie, ``The space-time phenomenology of
  photon absorbtion and inelastic electron scattering'',} \textit{ Annals
  Phys.} \textbf{ 72} (1972) 243,
  \href{http://dx.doi.org/10.1016/0003-4916(72)90242-4}{\doi{10.1016/0003-4916(72)90242-4}}.

\bibitem{Campbell:2015qma}
\hrefCMSnoop {}{J.~M. Campbell, R.~K. Ellis, and W.~T. Giele, ``A
  multi-threaded version of {MCFM}'',} \textit{ Eur. Phys. J. C} \textbf{ 75}
  (2015) 246,
  \href{http://dx.doi.org/10.1140/epjc/s10052-015-3461-2}{\doi{10.1140/epjc/s10052-015-3461-2}},
  \href{http://www.arXiv.org/abs/1503.06182}{\texttt{arXiv:1503.06182}}.

\bibitem{Gehrmann:2014fva}
T.~Gehrmann\hrefCMSnoop {}{ { et~al.}, ``{$W^+W^-$} production at hadron
  colliders in next to next to leading order {QCD}'',} \textit{ Phys. Rev.
  Lett.} \textbf{ 113} (2014) 212001,
  \href{http://dx.doi.org/10.1103/PhysRevLett.113.212001}{\doi{10.1103/PhysRevLett.113.212001}},
  \href{http://www.arXiv.org/abs/1408.5243}{\texttt{arXiv:1408.5243}}.

\bibitem{Czakon:2011xx}
\hrefCMSnoop {}{M.~Czakon and A.~Mitov, ``{Top++}: A program for the
  calculation of the top-pair cross-section at hadron colliders'',} \textit{
  Comput. Phys. Commun.} \textbf{ 185} (2014) 2930,
  \href{http://dx.doi.org/10.1016/j.cpc.2014.06.021}{\doi{10.1016/j.cpc.2014.06.021}},
  \href{http://www.arXiv.org/abs/1112.5675}{\texttt{arXiv:1112.5675}}.

\bibitem{Crivellin:2021ejk}
\hrefCMSnoop {}{A.~Crivellin and L.~Schnell, ``Complete {Lagrangian} and set of
  {Feynman} rules for scalar leptoquarks'',} \textit{ Comput. Phys. Commun.}
  \textbf{ 271} (2022) 108188,
  \href{http://dx.doi.org/10.1016/j.cpc.2021.108188}{\doi{10.1016/j.cpc.2021.108188}},
  \href{http://www.arXiv.org/abs/2105.04844}{\texttt{arXiv:2105.04844}}.

\bibitem{GEANT4:2002zbu}
\hrefCMSnoop {}{{{GEANT4}} Collaboration, ``{GEANT4 -- a simulation
  toolkit}'',} \textit{ Nucl. Instrum. Meth. A} \textbf{ 506} (2003) 250,
  \href{http://dx.doi.org/10.1016/S0168-9002(03)01368-8}{\doi{10.1016/S0168-9002(03)01368-8}}.

\bibitem{CMS:2019ied}
\hrefCMSnoop {}{{CMS Collaboration}, ``Performance of the reconstruction and
  identification of high-momentum muons in proton-proton collisions at
  {$\sqrt{s} =$ 13 TeV}'',} \textit{ JINST} \textbf{ 15} (2020) P02027,
  \href{http://dx.doi.org/10.1088/1748-0221/15/02/P02027}{\doi{10.1088/1748-0221/15/02/P02027}},
  \href{http://www.arXiv.org/abs/1912.03516}{\texttt{arXiv:1912.03516}}.

\bibitem{CMS:2018mdl}
\hrefCMSnoop {}{{CMS Collaboration}, ``Measurement of the differential
  {Drell-Yan} cross section in proton-proton collisions at {$ \sqrt{s} $ = 13
  TeV}'',} \textit{ JHEP} \textbf{ 12} (2019) 059,
  \href{http://dx.doi.org/10.1007/JHEP12(2019)059}{\doi{10.1007/JHEP12(2019)059}},
  \href{http://www.arXiv.org/abs/1812.10529}{\texttt{arXiv:1812.10529}}.

\bibitem{CMS:2014lcz}
\hrefCMSnoop {}{{CMS Collaboration}, ``Search for physics beyond the standard
  model in dilepton mass spectra in proton-proton collisions at {$ \sqrt{s}=8 $
  TeV}'',} \textit{ JHEP} \textbf{ 04} (2015) 025,
  \href{http://dx.doi.org/10.1007/JHEP04(2015)025}{\doi{10.1007/JHEP04(2015)025}},
  \href{http://www.arXiv.org/abs/1412.6302}{\texttt{arXiv:1412.6302}}.

\bibitem{ParticleDataGroup:2024cfk}
\hrefCMSnoop {}{{Particle Data Group} Collaboration, ``{Review of particle
  physics}'',} \textit{ Phys. Rev. D} \textbf{ 110} (2024) 030001,
  \href{http://dx.doi.org/10.1103/PhysRevD.110.030001}{\doi{10.1103/PhysRevD.110.030001}}.

\bibitem{Sirunyan:2019eyu}
\hrefCMSnoop {}{{CMS Collaboration}, ``Measurement of the top quark
  forward-backward production asymmetry and the anomalous chromoelectric and
  chromomagnetic moments in pp collisions at $ \sqrt{s} $ = 13 {TeV}'',}
  \textit{ JHEP} \textbf{ 06} (2020) 146,
  \href{http://dx.doi.org/10.1007/JHEP06(2020)146}{\doi{10.1007/JHEP06(2020)146}},
  \href{http://www.arXiv.org/abs/1912.09540}{\texttt{arXiv:1912.09540}}.

\bibitem{Manohar:2016nzj}
\hrefCMSnoop {}{A.~Manohar, P.~Nason, G.~P. Salam, and G.~Zanderighi, ``How
  bright is the proton? {A} precise determination of the photon parton
  distribution function'',} \textit{ Phys. Rev. Lett.} \textbf{ 117} (2016)
  242002,
  \href{http://dx.doi.org/10.1103/PhysRevLett.117.242002}{\doi{10.1103/PhysRevLett.117.242002}},
  \href{http://www.arXiv.org/abs/1607.04266}{\texttt{arXiv:1607.04266}}.

\bibitem{Manohar:2017eqh}
\hrefCMSnoop {}{A.~V. Manohar, P.~Nason, G.~P. Salam, and G.~Zanderighi, ``The
  photon content of the proton'',} \textit{ JHEP} \textbf{ 12} (2017) 046,
  \href{http://dx.doi.org/10.1007/JHEP12(2017)046}{\doi{10.1007/JHEP12(2017)046}},
\href{http://www.arXiv.org/abs/1708.01256}{\texttt{arXiv:1708.01256}}.

\bibitem{Conway:2011in}
\hrefCMSnoop {}{J.~S. Conway, ``Incorporating nuisance parameters in
  likelihoods for multisource spectra'',} in \textit{ {PHYSTAT 2011}}, p.~115.
\newblock 2011.
\newblock \href{http://www.arXiv.org/abs/1103.0354}{\texttt{arXiv:1103.0354}}.
\newblock
  \href{http://dx.doi.org/10.5170/CERN-2011-006.115}{\doi{10.5170/CERN-2011-006.115}}.

\bibitem{CMS-LUM-17-003}
\hrefCMSnoop {}{{CMS Collaboration}, ``Precision luminosity measurement in
  proton-proton collisions at $\sqrt{s} =$ 13 {TeV} in 2015 and 2016 at
  {CMS}'',} \textit{ Eur. Phys. J. C} \textbf{ 81} (2021) 800,
  \href{http://dx.doi.org/10.1140/epjc/s10052-021-09538-2}{\doi{10.1140/epjc/s10052-021-09538-2}},
  \href{http://www.arXiv.org/abs/2104.01927}{\texttt{arXiv:2104.01927}}.

\bibitem{CMS-PAS-LUM-17-004}
\href {https://cds.cern.ch/record/2621960/}{{CMS Collaboration}, ``{CMS}
  luminosity measurement for the 2017 data-taking period at $\sqrt{s}$ = 13
  {TeV}'',} CMS Physics Analysis Summary CMS-PAS-LUM-17-004, 2018.

\bibitem{CMS-PAS-LUM-18-002}
\href {https://cds.cern.ch/record/2676164/}{{CMS Collaboration}, ``{CMS}
  luminosity measurement for the 2018 data-taking period at $\sqrt{s}$ = 13
  {TeV}'',} CMS Physics Analysis Summary CMS-PAS-LUM-18-002, 2019.

\bibitem{CMS:2024onh}
\hrefCMSnoop {}{{CMS Collaboration}, ``The {CMS} statistical analysis and
  combination tool: {\textsc{Combine}}'',} \textit{ Comput. Softw. Big Sci.}
  \textbf{ 8} (2024) 19,
  \href{http://dx.doi.org/10.1007/s41781-024-00121-4}{\doi{10.1007/s41781-024-00121-4}},
  \href{http://www.arXiv.org/abs/2404.06614}{\texttt{arXiv:2404.06614}}.

\bibitem{FC}
\hrefCMSnoop {}{G.~J. Feldman and R.~D. Cousins, ``Unified approach to the
  classical statistical analysis of small signals'',} \textit{ Phys. Rev. D}
  \textbf{ 57} (1998) 3873,
  \href{http://dx.doi.org/10.1103/PhysRevD.57.3873}{\doi{10.1103/PhysRevD.57.3873}},
  \href{http://www.arXiv.org/abs/physics/9711021}{\texttt{arXiv:physics/9711021}}.

\bibitem{CLS1}
\hrefCMSnoop {}{T.~Junk, ``Confidence level computation for combining searches
  with small statistics'',} \textit{ Nucl. Instrum. Meth. A} \textbf{ 434}
  (1999) 435,
  \href{http://dx.doi.org/10.1016/S0168-9002(99)00498-2}{\doi{10.1016/S0168-9002(99)00498-2}},
\href{http://www.arXiv.org/abs/hep-ex/9902006}{\texttt{arXiv:hep-ex/9902006}}.

\bibitem{CLS2}
\hrefCMSnoop {}{A.~L. Read, ``Presentation of search results: The
  {CL$_{\text{s}}$} technique'',} \textit{ J. Phys. G} \textbf{ 28} (2002)
  2693,
\href{http://dx.doi.org/10.1088/0954-3899/28/10/313}{\doi{10.1088/0954-3899/28/10/313}}.

\bibitem{Cowan:2010js}
\hrefCMSnoop {}{G.~Cowan, K.~Cranmer, E.~Gross, and O.~Vitells, ``Asymptotic
  formulae for likelihood-based tests of new physics'',} \textit{ Eur. Phys. J.
  C} \textbf{ 71} (2011) 1554,
  \href{http://dx.doi.org/10.1140/epjc/s10052-011-1554-0}{\doi{10.1140/epjc/s10052-011-1554-0}},
  \href{http://www.arXiv.org/abs/1007.1727}{\texttt{arXiv:1007.1727}}.
  [Erratum: \DOI{10.1140/epjc/s10052-011-1554-0}].

\end{thebibliography}\endgroup
\cleardoublepage \appendix\section{The CMS Collaboration \label{app:collab}}\begin{sloppypar}\hyphenpenalty=5000\widowpenalty=500\clubpenalty=5000
\cmsinstitute{Yerevan Physics Institute, Yerevan, Armenia}
{\tolerance=6000
A.~Hayrapetyan, A.~Tumasyan\cmsAuthorMark{1}\cmsorcid{0009-0000-0684-6742}
\par}
\cmsinstitute{Institut f\"{u}r Hochenergiephysik, Vienna, Austria}
{\tolerance=6000
W.~Adam\cmsorcid{0000-0001-9099-4341}, J.W.~Andrejkovic, L.~Benato\cmsorcid{0000-0001-5135-7489}, T.~Bergauer\cmsorcid{0000-0002-5786-0293}, S.~Chatterjee\cmsorcid{0000-0003-2660-0349}, K.~Damanakis\cmsorcid{0000-0001-5389-2872}, M.~Dragicevic\cmsorcid{0000-0003-1967-6783}, P.S.~Hussain\cmsorcid{0000-0002-4825-5278}, M.~Jeitler\cmsAuthorMark{2}\cmsorcid{0000-0002-5141-9560}, N.~Krammer\cmsorcid{0000-0002-0548-0985}, A.~Li\cmsorcid{0000-0002-4547-116X}, D.~Liko\cmsorcid{0000-0002-3380-473X}, I.~Mikulec\cmsorcid{0000-0003-0385-2746}, J.~Schieck\cmsAuthorMark{2}\cmsorcid{0000-0002-1058-8093}, R.~Sch\"{o}fbeck\cmsAuthorMark{2}\cmsorcid{0000-0002-2332-8784}, D.~Schwarz\cmsorcid{0000-0002-3821-7331}, M.~Sonawane\cmsorcid{0000-0003-0510-7010}, W.~Waltenberger\cmsorcid{0000-0002-6215-7228}, C.-E.~Wulz\cmsAuthorMark{2}\cmsorcid{0000-0001-9226-5812}
% ...
\par}
\vskip\cmsinstskip
\dag:~Deceased\\
$^{1}$Also at Yerevan State University, Yerevan, Armenia\\
$^{2}$Also at TU Wien, Vienna, Austria\\
% ...
\end{sloppypar}
\end{document}